\documentclass[12pt]{iopart}
\usepackage{bm}
\usepackage{iopams}
\newcommand{\veps}{\varepsilon}

\newcommand{\dd}{\mbox{d}}
\newcommand{\thpint}{\int_{-\infty}^{\infty} \dd \, p \, \int_0^{2\pi} \dd \, \theta \,}
\newcommand{\pint}{\int_{-\infty}^{\infty} \dd \, p \,}
\begin{document}
\title[Stability of inhomogeneous quasi-stationary states]{Dynamical stability criterion for
inhomogeneous quasi-stationary states in long-range systems}
\author{Alessandro Campa$^1$ and Pierre-Henri Chavanis$^2$}
\address{$^1$ Complex Systems and Theoretical Physics Unit, Health and Technology Department,
Istituto Superiore di Sanit\`a, and INFN Roma1, Gruppo Collegato Sanit\`a,
00161 Roma, Italy}
\address{$^2$ Laboratoire de Physique Th\'eorique (IRSAMC), CNRS and UPS,
 Universit\'e de Toulouse, F-31062 Toulouse, France}
\ead{\mailto{campa@iss.infn.it}, \mailto{chavanis@irsamc.ups-tlse.fr}}

\begin{abstract}
We derive a necessary and sufficient condition of linear dynamical
stability for inhomogeneous Vlasov stationary states of the Hamiltonian
Mean Field (HMF) model. The condition is expressed by an explicit
disequality that has to be satisfied by the stationary state, and it generalizes
the known disequality for homogeneous stationary states. In addition, we
derive analogous disequalities that express necessary and
sufficient conditions of formal stability for the stationary
states. Their usefulness, from the point of view of linear dynamical
stability, is that they are simpler, although they provide only
sufficient criteria of linear stability. We show that for
homogeneous stationary states the relations become equal, and
therefore linear dynamical stability and formal stability become
equivalent.
\end{abstract}

\submitto{Journal of Statistical Mechanics: Theory and Experiment}
\maketitle

\section{Introduction}
\label{intro}

There are numerous distinctive features that characterize the
behaviour of many-body systems with long-range interactions, features
that are not present in systems with short-range interactions. These
peculiarities concern both the equilibrium properties, such as the
inequivalence of ensembles and negative specific heats in the
microcanonical ensemble, and the out-of-equilibrium dynamical
behaviour, such as the existence of long-lived quasi-stationary states
and of out-of-equilibrium phase transitions. The study of these
properties is interesting in its own, but it is also justified by the
many different physical systems in which long-range interactions play
the prominent role, e.g., self-gravitating systems
\cite{binney,padma,ijmpb}, unscreened Coulomb systems \cite{brydges}, some
models in plasma physics \cite{elskesc} and in hydrodynamics
\cite{chasomrob}, and trapped charged particles \cite{dubin}. Recent reviews give the state of the
art of the subject \cite{physrep,houch2008}.

In this paper, we treat the subject of the long-lived quasi-stationary
states (QSS). They are out-of-equilibrium states in which the
distribution functions are non-Boltzmannian, and their lifetime
increases with the size of the system as given by the number $N$ of
degrees of freedom; this increase generally scales as a power law in
$N$ but it can also be exponential. It has to be emphasized that the
QSS's are not related to the usual metastable states that are found also
in short-range systems. The latter are realized by local extrema of
thermodynamical potential (e.g., they are local maxima of the entropy
or local minima of the free energy, if these quantities are computed
as a function of an order parameter of the system), in which the
system is trapped until it is driven away by some perturbation, and
then heads towards the global extremum, i.e., the equilibrium
state. Global and local extrema, i.e., equilibrium and metastable
states, are obtained on the basis of the usual Boltzmann-Gibbs
statistics. The evaluation of these states can be done following
different routes; e.g., one can compute the partition function of the
$N$-body system or work at the level of the one-particle distribution
function, but only with a direct relation to the Boltzmann-Gibbs
statistics, that governs equilibrium.

On the other hand, QSS's in principle have nothing to do with some sort
of equilibrium state, global or local, of the system; nevertheless the
system can be trapped for macroscopic times in these states. We want
to underline that this fact does not imply at all any failure of the
Boltzmann-Gibbs description of the equilibrium states. It simply means
that the approach to equilibrium, that both in long-range and
short-range systems should be described, at least approximately, by a
kinetic equation, happens often in a manner, when long-range
interactions are present, in which the system resides for macroscopic
times in dynamical states that are very far from the equilibrium
states.

From this it should be clear that the ultimate reason for the
existence of the QSS's should be looked into dynamics, i.e., into the
representation of the dynamics via a kinetic equation. It turns out
that the dynamics of many-body long-range systems can be described
with a very good approximation, in a certain time range, by the Vlasov
equation for the one-particle distribution function. This equation is
also sometimes called the collisionless Boltzmann equation, since it
represents the interactions between the particles (whatever they might
be, they can even be stars or galaxies in astrophysical problems) by a
mean-field term, i.e., the two-body interaction potential averaged
over the whole system. The theoretical justification for this fact can
be looked upon at different levels of mathematical rigour. It is not
our task here to give this justification, but we think it is useful to
give a flavour of the reason from a physical point of view.

If a system is initially prepared in a state away from equilibrium, it
will evolve towards the equilibrium state because of the interactions
between the particles. In short-range systems each particle interacts
only with the nearby particles, and therefore the dynamical evolution
is determined by the ``collisions'' of any given particle with the few
others surrounding it.  Since the position of nearby particles are
strongly correlated, it is completely useless to approximate the field
acting on a particle by an averaged field, but it is necessary to find
a way to describe, in a kinetic equation, the collisions between close
particles. In long-range systems, even in the cases where the field at
close distances is strong, the field acting on a particular particle
is determined by all the others, and it looks quite plausible that an
averaged field can be a good approximation. Obviously, at the end, the
``collisional'' regime will take its toll and the Boltzmann-Gibbs
equilibrium will be realized, but before that point, the dynamics, to
a high degree of approximation, will follow the evolution determined
by the Vlasov equation.

It is then not surprising that a considerable amount of work has been
dedicated to the properties of the Vlasov equation in relation with
important systems like self-gravitating and plasma systems. It is
worth noting a difference between these two classes of systems.  In
fact, globally neutral plasmas have a spatially homogeneous equilibium
state, and it is in reference to this case that the theory of the
Vlasov linear stability of small perturbations of the homogeneous
equilibrium state has been developed \cite{case,balescu,nichols}. On the
contrary, the necessarily inhomogeneous states of large but finite
self-gravitating systems has motivated the research on the stability
properties of inhomogeneous stationary states of the Vlasov equation
\cite{binney,antonov62,lynd1967,bart,kandrup90,kandrup91}.

The Hamiltonian Mean Field (HMF) model \cite{konishi,antruf}, a simple
1D toy model of systems with long-range interactions, has been very
useful to study the various statistical and dynamical properties of
long-range systems. Also, a good amount of work has been dedicated
recently to the stability of Vlasov stationary states. The comments of
the previous paragraph suggest that one should be interested in the
stability of both homogeneous and inhomogeneous states. However,
mainly homogeneous states have been considered (both theoretically and
numerically, see, e.g.,
Refs. \cite{antruf,ik,latora,choi,yama,cvb,campa2008,cd}), and there exists
only few theoretical results for inhomogeneous states (see Secs. 3.3 and 4.4 of
\cite{cvb}, Appendix F of \cite{cd}, and \cite{cc} for polytropes).  
In this paper, we present a criterion for the Vlasov stability of
inhomogeneous stationary states of the HMF model that generalizes the
known criterion valid for homogeneous states.  The results presented
are based on the stability conditions for Vlasov stationary states
that have been derived in the (essentially astrophysical)
literature. We rederive here the results, and extend them, leaving the
interaction potential unspecified. We will introduce the HMF
interaction potential when the mentioned results will be used to
obtain explicit necessary and sufficient conditions on the stationary
distribution function.

We emphasize that the stability conditions already in the literature
are in the form of relations that have to be satisfied by the
perturbations to the stationary states.  To our knowledge, such
conditions have not been yet transformed in explicit conditions that
have to be obeyed by the stationary distribution itself.  Therefore,
such explicit conditions are the core of this paper.

It turns out that the necessary and sufficient conditions of linear
stability can be simplified at the price of obtaining conditions that
are only sufficient. This can be done in the framework of the formal
stability of the Vlasov stationary states. Therefore, for more
completeness and for a useful comparison, we find it useful to treat
also the problem of formal stability.

Section \ref{vlasovequa} introduces the stationary states in the
framework of the Vlasov equation. Sections from \ref{formalgeneral} to
\ref{formaltot} derive results later employed for the study of
linear stability and formal stability. This is done, for the HMF
model, in Sections \ref{stabhmftot} and \ref{formaltothmf}. Section
\ref{discuss} contains the conclusions.

\section{The Vlasov equation and the quasi-stationary states}
\label{vlasovequa}

This paper will use the HMF model as a benchmark for our analysis, and
therefore it is convenient to introduce the Vlasov equation from the
beginning in this framework. Consider $N$ particles of unit
mass moving on a circle, with the Hamiltonian of the system given by:
\begin{equation}
\label{hamilgen}
H = \frac{1}{2} \sum_{i=1}^N \frac{p_i^2}{2} + \frac{1}{2N}\sum_{i\ne j} V(\theta_i
-\theta_j) \, ,
\end{equation}
where $\theta_i \in [0,2\pi]$ is the angle giving the position of a
particle on the circle and $-\infty < p_i < \infty$ is its linear
momentum (equal to the velocity since the mass is unitary). The
$\frac{1}{N}$ normalization of the interaction potential $V(\theta_i
-\theta_j)$ is the usual one introduced in order to have an extensive
energy; it is equivalent to a system-size-dependent rescaling of time,
and it does not affect the study of the properties of the system. The
Vlasov equation associated to this system, governing the evolution of
the one-particle distribution function $f(\theta,p,t)$ is:
\begin{equation}
\label{vlasovgen}
\frac{\partial f(\theta,p,t)}{\partial t} + p \frac{\partial f(\theta,p,t)}{\partial \theta}
-\frac{\partial \Phi(\theta,t;f)}{\partial \theta}\frac{\partial f(\theta,p,t)}{\partial p} = 0 \, ,
\end{equation}
where $\Phi(\theta,t;f)$ is the mean field potential:
\begin{equation}
\label{meanfieldpotgen}
\Phi(\theta,t;f) = \int_{-\infty}^{\infty} \dd \, p' \, \int_0^{2\pi} \dd \, \theta' \,
V(\theta - \theta') f(\theta',p',t) \, .
\end{equation}
The last equation shows that the Vlasov equation (\ref{vlasovgen}) is
a nonlinear integrodifferential equation. It is immediate to see that
it conserves the normalization of $f(\theta,p,t)$:
\begin{equation}
\label{normgen}
\thpint f(\theta,p,t) = 1 \, ,
\end{equation}
and the total energy, given by:
\begin{equation}
\label{energgen}
E = \thpint  \left( \frac{p^2}{2} + \frac{1}{2}\Phi(\theta,t;f)\right)f(\theta,p,t) \, .
\end{equation}
The HMF model is obtained when $V(\theta - \theta')$ is a cosine
potential. In the following, we will present the criteria that
determine the dynamical stability of stationary states of the Vlasov equation
(\ref{vlasovgen}) for a general potential $V(\theta -\theta')$. These
criteria can be trivially generalized to arbitrary space
dimension. Afterwards, we will specialize to the HMF model when the
relations will be transformed in explicit conditions for the
stationary states.

As we have described in the Introduction, there is a time regime in
which the dynamics of the $N$-body system is, to a high degree of
accuracy, represented by the time evolution of the one-particle
distribution function as governed by the Vlasov equation. In this
framework, it is natural to expect that the QSS's will be associated to
stationary states (i.e., time independent states) of this equation. Of
course, these stationary states should also be stable, i.e., a small
perturbation should not drive the system away from the stationary
state. Then, it is natural to be interested in the dynamical stability
of the stationary states of the Vlasov equation.  It is easy to see
that any function of the form $f(\theta,p) = f(\frac{p^2}{2} +
\Phi(\theta;f))$ is stationary; therefore, in principle one is
interested in determining the stability of any such function.

The function $f(\frac{p^2}{2} + \Phi(\theta;f))$ is by definition
linearly stable if it is possible to choose the norm of the
perturbation $\delta f(\theta,p,t)$ at time $t=0$ such that this norm
remains smaller than any (small) positive number, provided that the
dynamics of the perturbation is governed by the linearized Vlasov
equation, with the linearization made around the stationary state.
Introducing the individual energy:
\begin{equation}
\label{enelocvlas}
\veps(\theta,p) \equiv \frac{p^2}{2} + \Phi(\theta;f) \, ,
\end{equation}
the linearized Vlasov equation for $\delta f(\theta,p,t)$ is easily obtained as:
\begin{equation}
\label{linearvlas}
\fl
\frac{\partial \delta f(\theta,p,t)}{\partial t} + p \frac{\partial \delta f(\theta,p,t)}
{\partial \theta} -\frac{\dd \Phi(\theta;f)}{\dd \theta}\frac{\partial \delta f(\theta,p,t)}
{\partial p} -\frac{\partial \Phi(\theta,t;\delta f)}{\partial \theta}
p f'(\veps(\theta,p)) = 0 \, ,
\end{equation}
where the potential $\Phi(\theta;f)$ is constant in time (and thus the
partial derivative with respect to $\theta$ has become a total
derivative) and where in the last term the functional dependence of
$f$ on $(\theta,p)$ only through $\veps(\theta,p)$ has been
exploited. The problem of the linear stability
associated to this equation has been treated long ago by Antonov
\cite{antonov62} in astrophysics. In order to have a self-contained
presentation, in section \ref{dynstabgeneral} we will reproduce,
although in a somewhat different formulation and scope, the Antonov
results, that later will be used to derive the stability conditions on
the stationary state $f(\frac{p^2}{2} + \Phi(\theta;f))$. These will
be necessary and sufficient conditions of linear stability.

As we have previously underlined, we are also interested in less
refined stability criteria, that provide only sufficient, but simpler,
conditions of linear stability. To that purpose, we can use the notion
of formal stability of a stationary point of a general dynamical
system. The stationary point is said to be formally stable
\cite{holmmars} if a conserved quantity has an extremum at the
stationary point and if the second variation about the stationary
point is either positive definite or negative definite. It can be
shown \cite{holmmars} that formal stability implies linear stability;
therefore, proving that a stationary point is formally stable gives a
sufficient condition for linear stability. Formal stability is also a
pre-requisit for nonlinear stability, although formal stability does
not imply nonlinear stability for infinite dimensional systems. With a
slight extension of definition, we will consider the formal stability
of a stationary point also for the cases in which both the
extremization of the conserved quantity and the sign of its second
variation are studied under some constraints. The problem will be
related, as it will be clear, by the fact that if one finds that the
second variation has a definite sign for the unconstrained case, this
is sufficient to have the same definite sign also for the constrained
case. However, the converse is wrong and this is similar to the notion
of ensembles inequivalence in statistical physics \cite{ellisineq}.

To study the formal stability problem, we use a general result
obtained in Ref.
\cite{ipshorw}. Again for a self-contained presentation, we find it useful to briefly
reproduce it here. This is done in the following section, in slighlty more general
terms than those necessary for the linear Vlasov equation.

\section{The maximization of a class of functionals}
\label{formalgeneral}

Let ${\cal L}$ be the functional space of real differentiable functions $f(\theta,p)$ defined for
$0 \le \theta \le 2\pi$ and $p \in {\cal R}$. We also assume that the functions decay
sufficiently fast for $p \to \infty$ so that the integral of $p^2f$ is finite. The scalar
product in this space is naturally defined by:
\begin{equation}
\label{scaldef}
\langle g,f \rangle = \thpint g(\theta,p)f(\theta,p) \, .
\end{equation}
We consider here the problem \cite{ipshorw} of finding the constrained maximum of the
functional:
\begin{equation}
\label{expfunct}
S[f] = - \thpint C\left( f(\theta,p) \right) \, ,
\end{equation}
with the function $C(x)$ at least twice differentiable and strictly
convex, i.e., with the second derivative strictly positive. The
constraints are given by two functionals. The first is a
linear-quadratic expression:
\begin{eqnarray}
\label{enetotexam}
E &=& \thpint a(\theta,p) f(\theta,p) \nonumber \\ &+& \frac{1}{2}\thpint
\int_{-\infty}^{\infty} \dd \, p' \, \int_0^{2\pi} \dd \, \theta' \,
f(\theta,p) b(\theta,p,\theta',p')f(\theta',p') \, ,
\end{eqnarray}
that we can call the ``total energy'' (in analogy with
Eq. (\ref{energgen})), and that is constrained to have a given value
$E_0$. In this expression, $a$ and $b$ are two given functions,
with $b$ possessing the symmetry property
$b(\theta',p',\theta,p)=b(\theta,p,\theta',p')$. The second constraint
is the normalization:
\begin{equation}
\label{normexam}
I = \thpint f(\theta,p) = 1 \, .
\end{equation}
By using Lagrange multipliers $\beta$ and $\mu$ the extremum is given by
equating to zero the first order variation:
\begin{equation}
\label{extremfirst}
\delta S - \beta \delta E - \mu \delta I = 0 \, .
\end{equation}
We thus have:
\begin{equation}
\label{extremexam}
\fl
-C'\left( f(\theta,p) \right) - \beta \left[ a(\theta,p) +
\int_{-\infty}^{\infty} \dd \, p' \, \int_0^{2\pi} \dd \, \theta' \,
b(\theta,p,\theta',p') f(\theta',p') \right] - \mu = 0 \, .
\end{equation}
If we denote the ``individual energy'' (in analogy with Eq. (\ref{enelocvlas})) by:
\begin{equation}
\label{eneexam}
\veps(\theta,p) \equiv a(\theta,p) +
\int_{-\infty}^{\infty} \dd \, p' \, \int_0^{2\pi} \dd \, \theta' \,
b(\theta,p,\theta',p') f(\theta',p') \, ,
\end{equation}
then the extremum relation can be written as:
\begin{equation}
\label{extremexam1}
-C'\left( f(\theta,p) \right) = \beta \veps(\theta,p) + \mu \, .
\end{equation}
From the convexity property of $C(x)$ it follows that this relation can be
inverted to give:
\begin{equation}
\label{fextremexam}
f(\theta,p) = F (\beta \veps + \mu) \equiv f(\veps) \, ,
\end{equation}
where $F$ is the inverse function of $-C'$. Inserting this function in
Eqs. (\ref{enetotexam}) and (\ref{normexam}) we obtain the values of
the Lagrange multipliers. It is clear that Eq. (\ref{eneexam}) is also
a consistency equation.  We note that in order to interpret
$f(\theta,p)$ as a distribution function, acceptable functions $C(f)$
in Eq. (\ref{expfunct}) are only those that, through Eq. (\ref{fextremexam}),
provide a positive definite function. We also note the identity:
\begin{equation}
\label{id}
f'(\veps(\theta,p))=-\frac{\beta}{C''(f(\theta,p))},
\end{equation}
that is obtained by differentiating Eq. (\ref{extremexam1}). From the
convexity property of $C(x)$ it follows that
$\frac{1}{\beta}f'(\veps(\theta,p))$ is negative definite. On the
other hand, since we have $\frac{\partial f}{\partial p} = p
f'(\veps(\theta,p))$, the integrability in $p$ requires that, if
$f'(\veps(\theta,p))$ has a definite sign, this sign must be
negative. Therefore, $\beta$ is restricted to positive values. We
therefore conclude that the extremization of $S$ at fixed $E$ and $I$
determines distribution functions of the form
$f=f(\veps)$ with $f'(\veps)< 0$. 

The extremum so obtained will be a maximum only if the second order variation
of $S[f]$, for all the allowed displacements $\delta f(\theta,p)$, is
negative definite. The variation of the functional $S[f]$ is given,
up to second order, by:
\begin{equation}
\label{variatsec}
\fl
\delta S = - \thpint \left[ C'(f(\theta,p))\delta f(\theta,p) +
\frac{1}{2}C''(f(\theta,p))(\delta f(\theta,p))^2\right] \, ,
\end{equation}
with the derivatives computed at the extremal point. Then, from Eqs.
(\ref{extremexam1}), (\ref{fextremexam}) and (\ref{id}) we obtain:
\begin{equation}
\label{variatsec1}
\fl
\delta S = \thpint \left[(\beta \veps(\theta,p)+\mu)\delta f(\theta,p) +
\frac{1}{2}\frac{\beta}{f'(\veps(\theta,p))}
(\delta f(\theta,p))^2\right] \, .
\end{equation}
We can transform this expression by using the fact that the variations of $E$ and $I$
must identically vanish for the allowed displacements $\delta f(\theta,p)$.
These variations are given by:
\begin{eqnarray}
\label{variatE}
\fl
\delta E &=& \thpint \veps(\theta,p) \delta f(\theta,p) \nonumber \\ \fl &+&
\frac{1}{2} \thpint \int_{-\infty}^{\infty} \dd \, p' \, \int_0^{2\pi} \dd \,
\theta' \,
\delta f(\theta,p) b(\theta,p,\theta',p') \delta f(\theta',p') \equiv 0
\end{eqnarray}
and
\begin{equation}
\label{variatI}
\delta I = \thpint \delta f(\theta,p) \equiv 0
\end{equation}
respectively. Adding to Eq. (\ref{variatsec1}) the zero valued expression
$-\beta \delta E - \mu \delta I$ we have:
\begin{eqnarray}
\label{variatsec2}
\fl
\delta S &=& \frac{1}{2}\beta \thpint \frac{1}{f'(\veps(\theta,p))}
(\delta f(\theta,p))^2 \nonumber \\ \fl &-& 
\frac{1}{2} \beta \thpint \int_{-\infty}^{\infty} \dd \, p' \, \int_0^{2\pi} \dd \,
\theta' \,
\delta f(\theta,p) b(\theta,p,\theta',p') \delta f(\theta',p') \, .
\end{eqnarray}
The right-hand side of this expression must be negative definite for
all allowed displacements $\delta f(\theta,p)$, i.e., for all those
that at first order do not change $E$ and $I$. 

The problem of maximizing $S[f]$ at constant $E$ and $I$ can be shown to be equivalent
to that of minimizing the energy $E$ at constant $S$ and $I$.
In fact, using Lagrange multipliers $1/\beta$ and $-\mu/\beta$, the equation of the
first order variation is now:
\begin{equation}
\label{extremfirstene}
\delta E - \frac{1}{\beta} \delta S + \frac{\mu}{\beta} \delta I = 0 \, ,
\end{equation}
which is the same as Eq. (\ref{extremfirst}); then the solution is
again given by Eq. (\ref{extremexam}). Now, we have to study the
variation of $E$, that up to second order is given by the left-hand
side of Eq. (\ref{variatE}), i.e.:
\begin{eqnarray}
\label{variatEb}
\fl
\delta E &=& \thpint \veps(\theta,p) \delta f(\theta,p) \nonumber \\ \fl &+&
\frac{1}{2} \thpint \int_{-\infty}^{\infty} \dd \, p' \, \int_0^{2\pi} \dd \,
\theta' \,
\delta f(\theta,p) b(\theta,p,\theta',p') \delta f(\theta',p') \, .
\end{eqnarray}
The variations of $S$ and $I$ must identically vanish; the latter is
expressed by Eq. (\ref{variatI}), while the former is given by:
\begin{equation}
\label{variatS}
\fl
\delta S = \thpint \left[(\beta \veps(\theta,p)+\mu)\delta f(\theta,p) +
\frac{1}{2}\frac{\beta}{f'(\veps(\theta,p))}
(\delta f(\theta,p))^2\right] \equiv 0 \, .
\end{equation}
Adding to Eq. (\ref{extremfirstene}) the zero valued expression $-\frac{1}{\beta}
\delta S + \frac{\mu}{\beta} \delta I$ we arrive at:
\begin{eqnarray}
\label{variatEc}
\fl
\delta E &=& -\frac{1}{2} \thpint \frac{1}{f'(\veps(\theta,p))}
(\delta f(\theta,p))^2 \nonumber \\ \fl &+& 
\frac{1}{2} \thpint \int_{-\infty}^{\infty} \dd \, p' \, \int_0^{2\pi} \dd \,
\theta' \,
\delta f(\theta,p) b(\theta,p,\theta',p') \delta f(\theta',p') \, .
\end{eqnarray}
This is the same as Eq. (\ref{variatsec2}) divided by $-\beta <
0$. Therefore, Eq. (\ref{variatEc}) is positive definite, i.e., $E$ is
minimum at the stationary state, if Eq. (\ref{variatsec2}) is negative
definite.  To complete the proof of the equivalence between the
maximization of $S$ at constant $E$ and $I$ and the minimization of
$E$ at constant $S$ and $I$ we have to see that in both cases the
allowed displacements $\delta f(\theta,p)$ are the same. This can be
deduced in the following way. For Eq. (\ref{variatsec2}) the allowed
displacements are all those that at first order give $\delta E =
\delta I = 0$. By Eq.  (\ref{extremfirst}), they also at first order give
$\delta S = 0$; then, they are also allowed for Eq. (\ref{variatEc}). In turn,
the allowed displacements for Eq. (\ref{variatEc}) are all those that at
first order give $\delta S = \delta I = 0$. By
Eq. (\ref{extremfirstene}), they also at first order give $\delta E =
0$; then, they are also allowed for Eq. (\ref{variatsec2}). This
concludes the proof.

After treating the problem of the linear stability of the stationary states of the Vlasov
equation, we will use the results of this section to study their formal stability.

\section{The linear stability of Vlasov stationary states}
\label{dynstabgeneral}

The energy functional in the previous section, Eq. (\ref{enetotexam}), was more general than the
one associated to the Vlasov equation, Eq.  (\ref{energgen}). The former reduces to the latter
when the functions $a(\theta,p)$ and $b(\theta,p,\theta',p')$ are related to the kinetic energy
and to the potential energy of the system, respectively; namely, when
$a= \frac{p^2}{2}$ and $b = V(\theta - \theta')$. For convenience, we rewrite
here the relevant expressions. We have the individual energy
\begin{equation}
\label{enelocvlas1}
\veps(\theta,p) = \frac{p^2}{2} + \Phi(\theta;f) \, ,
\end{equation}
with the mean field potential
\begin{equation}
\label{meanfieldpotgen1}
\Phi(\theta;f) = \int_{-\infty}^{\infty} \dd \, p' \, \int_0^{2\pi} \dd \, \theta' \,
V(\theta - \theta') f(\theta',p') \, ,
\end{equation}
and the total energy
\begin{equation}
\label{energgen1}
E[f] = \thpint  \left( \frac{p^2}{2} + \frac{1}{2}\Phi(\theta,t;f)\right)f(\theta,p,t) \, .
\end{equation}
We will consider stationary states that are associated to the
extremization of functionals of the form (\ref{expfunct}). We have
seen that in this case $f(\theta,p)=f(\veps(\theta,p))$ with
$f'(\veps(\theta,p)) < 0$. For more compactness, we will use the
notation $\gamma(\theta,p)\equiv f'(\veps(\theta,p))$. As remarked above, we
follow and complete the treatment of Antonov
\cite{antonov62}.

In the following we will need the usual extension of the scalar product defined in
Eq. (\ref{scaldef}) to complex valued functions:
\begin{equation}
\label{scaldefc}
\langle g,f \rangle = \thpint g^*(\theta,p)f(\theta,p) \, ,
\end{equation}
where the asterisk denotes complex conjugation.

It is not difficult to see that the linearized Vlasov equation (\ref{linearvlas}), determining
the dynamics of $\delta f(\theta,p)$, can be cast in the form:
\begin{equation}
\label{dynamexam}
\frac{\partial \delta f}{\partial t} (\theta,p,t) = -\gamma(\theta,p)(DK\delta f)(\theta,p,t) \, ,
\end{equation}
where $D$ is the antisymmetric linear differential operator (advective
operator):
\begin{equation}
\label{defD}
(Dg)(\theta,p) = p\frac{\partial}{\partial \theta}g(\theta,p) -
\frac{\dd \Phi}{\dd \theta}\frac{\partial}{\partial p}g(\theta,p) \, ,
\end{equation}
while $K$ is the linear integral operator:
\begin{equation}
\label{defK}
(Kg)(\theta,p) = \frac{1}{\gamma(\theta,p)} g(\theta,p) - \Phi(\theta;g) \, .
\end{equation}
This can be obtained by seeing that $D\veps = 0$; then, the action of
$D$ on any function of $\veps$ gives zero; in particular
$D\gamma = 0$. Finally, it can be easily checked that the operator $B
\equiv DKD$, needed shortly, is hermitian.

The stationary point of the dynamics $\delta f(\theta,p) \equiv 0$
will be linearly stable iff: {\it (i)} all nonsecular solutions of the
type $\delta f(\theta,p,t) = \delta f(\theta,p,0) \exp (\lambda t)$
have eigenvalues $\lambda$ with non positive real part; {\it (ii)} in
the presence of secular terms (i.e., if there are eigenvalues with an
algebraic multiplicity larger than the geometric multiplicity), the
associated eigenvalue must have a negative real part.

Before proceeding further, we need to put in evidence the properties of the
operators $D$ and $K$ when acting on functions that are either
symmetric or antisymmetric in $p$. From the definition of $D$ in
Eq. (\ref{defD}) it is clear that $D$ transforms symmetric functions
in antisymmetric functions, and viceversa. Concerning $K$, we see from
its definition in Eq. (\ref{defK}) that it maintains the symmetry of
the functions; however, since for antisymmetric functions
$g_a(\theta,p)$ we have $\Phi(\theta;g_a)\equiv 0$, the action of
$K$ in this case simplifies in
\begin{equation}
\label{defKant}
(Kg_a)(\theta,p) = \frac{1}{\gamma(\theta,p)} g_a(\theta,p) \, .
\end{equation}

We now suppose that $\delta f(\theta,p;\lambda)$ is the eigenfunction associated to the
eigenvalue $\lambda$. We then have, from Eq. (\ref{dynamexam}):
\begin{equation}
\label{eigen1}
\lambda \delta f(\theta,p;\lambda) =- \gamma (DK \delta f)(\theta,p;\lambda) \, ,
\end{equation}
where for simplicity we have dropped the dependence of $\gamma$ on the coordinates.
We now separate $\delta f$ in the symmetric and antisymmetric parts:
$\delta f = \delta f_s + \delta f_a$. Taking into account the mentioned properties
of the operators $D$ and $K$ we obtain
\begin{equation}
\label{eigen1s}
\lambda \delta f_s(\theta,p;\lambda) = -(D \delta f_a)(\theta,p;\lambda)
\end{equation}
and
\begin{equation}
\label{eigen1a}
\lambda \delta  f_a(\theta,p;\lambda) = -\gamma (DK \delta f_s)(\theta,p;\lambda) \, .
\end{equation}
If we multiply the second of these equations by $\lambda$ and substitute
$\lambda \delta f_s$ from the first we have:
\begin{equation}
\label{eigen2}
\lambda^2 \delta f_a(\theta,p;\lambda) = \gamma (DKD \delta f_a)(\theta,p;\lambda) =
\gamma (B \delta f_a)(\theta,p;\lambda) \, .
\end{equation}
If $\lambda \ne 0$, Eqs. (\ref{eigen1s}) and (\ref{eigen1a}) prove two properties. The first
is that an eigenfunction cannot be either symmetric or antisymmetric, but both components
must be nonvanishing. The second is that, if $\delta f_s + \delta f_a$ is associated to the
eigenvalue $\lambda$, then $\delta f_s - \delta f_a$ is associated to the eigenvalue
$-\lambda$.

From Eq. (\ref{eigen2}) we have:
\begin{equation}
\label{eigen3}
\lambda^2 \frac{\delta f_a(\theta,p;\lambda)}{\gamma(\theta,p)} =
(B \delta f_a)(\theta,p;\lambda) \, .
\end{equation}
The scalar product of both sides of this expression with $\delta f_a$ gives:
\begin{equation}
\label{scallam1}
\lambda^2 \thpint \frac{|\delta f_a(\theta,p;\lambda)|^2}{\gamma(\theta,p)}
= \langle \delta f_a, B \delta f_a \rangle \, ,
\end{equation}
while the scalar product of its complex conjugate with $\delta f_a^*$ gives,
exploiting the hermiticity of $B$:
\begin{equation}
\label{scallam2}
(\lambda^*)^2 \thpint \frac{|\delta f_a(\theta,p;\lambda)|^2}{\gamma(\theta,p)}
= \langle \delta f_a, B \delta f_a \rangle \, .
\end{equation}
Since $\gamma$ is negative definite, Eqs. (\ref{scallam1}) and
(\ref{scallam2}) imply that, if $\lambda \ne 0$, $\lambda^2$ is
necessarily real, and therefore $\lambda$ is either real or pure
imaginary. In much the same way, it can be shown that, if $\delta
f(\theta,p;\lambda_1)$ and $\delta f(\theta,p;\lambda_2)$ correspond
to two different non zero eigenvalues, then:
\begin{equation}
\label{scallam2b}
\langle \delta f_a(\lambda_1), B \delta f_a(\lambda_2) \rangle = 0 \, .
\end{equation}
The case $\lambda = 0$ will be considered later.

Let us now assume that an eigenvalue $\lambda$ has an algebraic multiplicity larger
than its geometric multiplicity. Then, if $\delta f(\theta,p;\lambda)$ is an
eigenvector associated to this eigenvalue, there will also exist a solution of Eq.
(\ref{dynamexam}) given by $[t\delta f(\theta,p;\lambda) + \delta f^{(1)}(\theta,p;\lambda)]
\exp (\lambda t)$. Substituting in Eq. (\ref{dynamexam}) and using Eq. (\ref{eigen1})
we obtain:
\begin{equation}
\label{eigen4}
\delta f(\theta,p;\lambda) + \lambda \delta f^{(1)}(\theta,p;\lambda) =
-\gamma (DK \delta f^{(1)})(\theta,p;\lambda) \, .
\end{equation}
Separating both $\delta f$ and $\delta f^{(1)}$ in the symmetric and antisymmetric
parts we obtain:
\begin{equation}
\label{eigen4s}
\delta f_s(\theta,p;\lambda) + \lambda \delta f^{(1)}_s(\theta,p;\lambda) =
-(D \delta f^{(1)}_a)(\theta,p;\lambda)
\end{equation}
and
\begin{equation}
\label{eigen4a}
\delta f_a(\theta,p;\lambda) + \lambda \delta  f^{(1)}_a(\theta,p;\lambda) =-
\gamma (DK \delta f^{(1)}_s)(\theta,p;\lambda) \, .
\end{equation}
If we multiply the second of these equations by $\lambda$ and we substitute
$\lambda \delta f^{(1)}_s$ from the first we have:
\begin{equation}
\label{eigen5}
\fl
\lambda \delta f_a(\theta,p;\lambda) + \lambda^2 \delta f^{(1)}_a(\theta,p;\lambda) =
\gamma (B\delta f^{(1)}_a)(\theta,p;\lambda) +\gamma (DK \delta f_s)(\theta,p;\lambda) \, .
\end{equation}
Using Eq. (\ref{eigen1a}) we arrive at:
\begin{equation}
\label{eigen6}
\frac{2\lambda  \delta f_a(\theta,p;\lambda) + \lambda^2 \delta f^{(1)}_a(\theta,p;\lambda)}
{\gamma(\theta,p)} = (B \delta f^{(1)}_a)(\theta,p;\lambda) \, .
\end{equation}
The scalar product of this expression with $\delta f_a$ gives:
\begin{equation}
\label{scallam3}
\fl
\thpint \frac{2\lambda|\delta f_a(\theta,p;\lambda)|^2 + \lambda^2 \delta f_a^*(\theta,p;\lambda)
\delta f_a^{(1)}(\theta,p;\lambda)}{\gamma(\theta,p)}
= \langle \delta f_a, B \delta f^{(1)}_a \rangle \, .
\end{equation}
Substracting from this equation the one that is obtained by forming the scalar product
of the complex conjugate of Eq. (\ref{eigen3}) with $\delta f_a^{(1)*}$ we have:
\begin{equation}
\label{scallam4}
\lambda \thpint \frac{|\delta f_a(\theta,p;\lambda)|^2}{\gamma(\theta,p)} = 0 \, .
\end{equation}
If $\lambda \ne 0$ we deduce that $\delta f_a(\theta,p;\lambda) \equiv 0$, since $\gamma$
is negative definite; then also $\delta f_s(\theta,p;\lambda) \equiv 0$ and thus
$\delta f(\theta,p;\lambda) \equiv 0$. This shows that no eigenvalue $\lambda$ different
from $0$ has an algebraic multiplicity larger than its geometric multiplicity.

We now consider the case $\lambda = 0$.
We note that the presence of a zero eigenvalue with an algebraic multiplicity larger
than the geometrical multiplicity would imply the presence of a solution of the equation
of motion (\ref{dynamexam}) of the form $[t\delta f(\theta,p;0) +
\delta f^{(1)}(\theta,p;0)]$, and therefore the stationary point $\delta f(\theta,p) \equiv 0$
would be linearly unstable. In the following we need to consider also the possibility that
the difference between the algebraic and the geometric multiplicity of the zero eigenvalue
is such that solutions with higher powers of the time $t$ exist.

For $\lambda = 0$, Eqs. (\ref{eigen1s}) and (\ref{eigen1a}) become:
\begin{equation}
\label{eigen1s0}
(D \delta f_a)(\theta,p;0) = 0
\end{equation}
and
\begin{equation}
\label{eigen1a0}
(DK \delta f_s)(\theta,p;0) = 0 \, .
\end{equation}
We note that Eq. (\ref{eigen1s0}) implies $(B \delta f_a)(\theta,p;0) = 0$.

In the case the eigenvalue $\lambda = 0$ has an algebraic multiplicity
larger than the geometric multiplicity, we obtain, from Eqs. (\ref{eigen4s}) and
(\ref{eigen4a}):
\begin{equation}
\label{eigen4s0}
\delta f_s(\theta,p;0) = - (D \delta f^{(1)}_a)(\theta,p;0)
\end{equation}
and
\begin{equation}
\label{eigen4a0}
\delta f_a(\theta,p;0) = - \gamma (DK \delta f^{(1)}_s)(\theta,p;0) \, .
\end{equation}
Substitution of Eq. (\ref{eigen4s0}) in Eq. (\ref{eigen1a0}) gives $(B \delta f^{(1)}_a)
(\theta,p;0) = 0$. Dividing both sides of Eq. (\ref{eigen4a0}) by $\gamma$ and forming
the scalar product with $\delta f_a$ we get:
\begin{equation}
\label{scallam5}
\fl
\thpint \frac{|\delta f_a(\theta,p;0)|^2}{\gamma(\theta,p)} = 
-\langle \delta f_a, DK \delta f^{(1)}_s \rangle =
 \langle D\delta f_a, K \delta f^{(1)}_s \rangle = 0 \, ,
\end{equation}
where in the last step we have used Eq. (\ref{eigen1s0}). It follows that
$\delta f_a(\theta,p;0) \equiv 0$, and then $\delta f_s(\theta,p;0) \ne 0$ in order
to have a non trivial solution. Then, from Eq. (\ref{eigen4s0}) we obtain that
$\delta f^{(1)}_a(\theta,p;0) \ne 0$ and $(D\delta f^{(1)}_a)(\theta,p;0) =
-\delta f_s(\theta,p;0) \ne 0$.

If there exists a solution of the equation
of motion (\ref{dynamexam}) of the form $[t^2\delta f(\theta,p;0) +
t\delta f^{(1)}(\theta,p;0) + \delta f^{(2)}(\theta,p;0)]$, then the evaluation of
(\ref{dynamexam}) at $t=0$ gives:
\begin{equation}
\label{eigen70}
\delta f^{(1)}(\theta,p;0) =- \gamma (DK \delta f^{(2)})(\theta,p;0) \, .
\end{equation}
The usual separation in the symmetric and antisymmetric parts gives:
\begin{equation}
\label{eigen7s0}
\delta f^{(1)}_s(\theta,p;0) =- (D \delta f^{(2)}_a)(\theta,p;0)
\end{equation}
and
\begin{equation}
\label{eigen7a0}
\delta f^{(1)}_a(\theta,p;0) = -\gamma (DK \delta f^{(2)}_s)(\theta,p;0) \, .
\end{equation}
The substitution of Eq. (\ref{eigen7s0}) in Eq. (\ref{eigen4s0}), taking into account that
$\delta f_a(\theta,p;0) = 0$, gives $(B \delta f^{(2)}_a)(\theta,p;0) = 0$.
Finally, it can be shown that a solution of the form $[t^3\delta f(\theta,p;0) +
t^2\delta f^{(1)}(\theta,p;0) + t\delta f^{(2)}(\theta,p;0) + \delta f^{(3)}(\theta,p;0)]$
cannot exist. In fact, the evaluation of (\ref{dynamexam}) at $t=0$ gives, after
separation in the symmetric and antisymmetric parts:
\begin{equation}
\label{eigen8s0}
\delta f^{(2)}_s(\theta,p;0) = -(D \delta f^{(3)}_a)(\theta,p;0) \, .
\end{equation}
Substitution in Eq. (\ref{eigen7a0}) gives:
\begin{equation}
\label{eigen7a0b}
\delta f^{(1)}_a(\theta,p;0) = \gamma (B \delta f^{(3)}_a)(\theta,p;0) \, .
\end{equation}
Dividing both sides by $\gamma$ and forming the scalar product with $\delta f^{(1)}_a$
we obtain:
\begin{equation}
\label{scallam6}
\fl
\thpint \frac{|\delta f^{(1)}_a(\theta,p;0)|^2}{\gamma(\theta,p)} = 
\langle \delta f^{(1)}_a, B \delta f^{(3)}_a \rangle =
\langle B\delta f^{(1)}_a, \delta f^{(3)}_s \rangle = 0 \, ,
\end{equation}
since $(B \delta f^{(1)}_a)(\theta,p;0) = 0$. This implies that
$\delta f^{(1)}_a(\theta,p;0) = 0$, that in turns gives, from Eq. (\ref{eigen4s0}),
$\delta f_s(\theta,p;0)=0$. This is not acceptable, since we already have
$\delta f_a(\theta,p;0) = 0$.

Summarizing all the results, we have that the initial value of $\delta
f(\theta,p)$ can be decomposed in general as:
\begin{eqnarray}
\label{decompf}
\delta f(\theta,p) &=& \sum_{\lambda \ne 0} c(\lambda) \delta f(\theta,p;\lambda) \nonumber \\
&+& \sum_{j=1}^r \left[ c_j \delta f_j(\theta,p;0) + c^{(1)}_j \delta f^{(1)}_j(\theta,p;0)
+ c^{(2)}_j \delta f^{(2)}_j(\theta,p;0)\right] \, ,
\end{eqnarray}
where the sum over the nonzero eigenvalues stands also for an integral
if the eigenvalues are continuously distributed. The sum over $r$
different contributions coming from the zero eigenvalue takes into
account possible separation into disjoint eigenspaces; some of the
corresponding functions might have only $\delta f_j(\theta,p;0)$
different from zero, and some only $\delta f_j(\theta,p;0)$ and
$\delta f^{(1)}_j(\theta,p;0)$.  The fact that $(B \delta
f_{j,a})(\theta,p;0) = (B \delta f^{(1)}_{j,a})(\theta,p;0) = (B
\delta f^{(2)}_{j,a})(\theta,p;0) = 0$ implies that if we take the
antisymmetric part of $\delta f(\theta,p)$ in Eq. (\ref{decompf}) and
form the scalar product $\langle
\delta f_a, B \delta f_a \rangle$, only the eigenfunctions corresponding to the
eigenvalues $\lambda$ different from zero contribute. Precisely, using the
orthogonality property (\ref{scallam2b}) we have:
\begin{equation}
\label{scallam7}
\langle \delta f_a, B\delta f_a \rangle = \sum_{\lambda \ne 0} |c(\lambda)|^2
\langle \delta f_a(\lambda), B\delta f_a(\lambda) \rangle \, .
\end{equation}
With the use of Eq. (\ref{scallam2}) we get:
\begin{equation}
\label{scallam8}
\langle \delta f_a, B\delta f_a \rangle = \sum_{\lambda \ne 0} |c(\lambda)|^2
\lambda^2 \thpint \frac{|\delta f_a(\theta,p;\lambda)|^2}{\gamma(\theta,p)}.
\end{equation}
This is the first important expression of this section. From it we deduce that,
if the stationary point $\delta f(\theta,p) = 0$ is linearly stable,
then necessarily the last expression is positive for all cases in which
not all the coefficients $c(\lambda)$ are zero. In fact, the linear stability requires
that all eigenvalues different from zero are pure imaginary (we recall that
if a real negative eigenvalue exists, also its opposite exists and leads to instability;
we also recall
that the antisymmetric part of any eigenfunction corresponding to an eigenvalue
different from zero is nonvanishing). From this and from
the fact that $\gamma$ is negative definite, our statement follows.

The scalar product (\ref{scallam8}) can be zero for a nonvanishing
$\delta f$ if one or some of the coefficients $c_j$, $c^{(1)}_j$,
$c^{(2)}_j$ are non zero. We know that, if one of the eigenspaces,
e.g. the one corresponding to $j=j_0$ in Eq.  (\ref{decompf}), has an
algebraic multiplicity larger than its geometric multiplicity, the
stationary point $\delta f(\theta,p) = 0$ is linearly unstable. In
this case, taking $\delta f(\theta,p) = \delta
f_{j_0}^{(1)}(\theta,p)$ we have $\langle \delta f_a, B\delta f_a
\rangle = 0$ with, according to what was proved just after
Eq. (\ref{scallam5}), $(D \delta f_a)(\theta,p) \ne 0$. On the other
hand, if all the eigenspaces corresponding to $\lambda = 0$ have equal
algebraic and geometric multiplicities, then only the terms with $c_j$
appear in Eq. (\ref{decompf}) as the contribution from the zero
eigenvalue. But in this case, it follows from Eq. (\ref{eigen1s0}) that the
scalar product (\ref{scallam8}) can be zero only for a function
$\delta f(\theta,p)$ such that $(D \delta f_a)(\theta,p) = 0$.

In conclusion, the necessary and sufficient condition for the linear stability
is that the scalar product (\ref{scallam8}) is nonnegative, and it is zero
only for functions $\delta f(\theta,p)$ such that $(D \delta f_a)(\theta,p) = 0$.

We now rewrite the scalar product $\langle \delta f_a, B \delta f_a \rangle$ in
another way. Precisely:
\begin{eqnarray}
\label{transfscal}
\langle \delta f_a, B \delta f_a \rangle &=& \langle \delta f_a, DKD \delta f_a \rangle
= - \langle D \delta f_a, KD \delta f_a \rangle \nonumber \\ &=&
-\langle D \delta f_a, \frac{1}{\gamma} D \delta f_a \rangle +
\langle D \delta f_a, \Phi (D\delta f_a) \rangle \, ,
\end{eqnarray}
where use has been made of the definition of the operator $K$ in Eq. (\ref{defK}).
Using the definition of $\Phi$ in Eq. (\ref{meanfieldpotgen}) we finally obtain:
\begin{eqnarray}
\label{scallam9}
\fl
\langle \delta f_a, B \delta f_a \rangle &=& -
\thpint \frac{1}{\gamma(\theta,p)}|(D \delta f_a)(\theta,p)|^2 \\ \fl &+& 
\thpint \int_{-\infty}^{\infty} \dd \, p' \, \int_0^{2\pi} \dd \, \theta' \,
(D \delta f_a)^*(\theta,p) V(\theta -\theta') (D \delta f_a)(\theta',p') \nonumber \, .
\end{eqnarray}
We should note two things. Firstly, since our equation of motion is real,
it is always possible to choose $\delta f(\theta,p)$ real. Secondly, it is not
difficult to check, form the last expression, that if we consider the whole
function $\delta f(\theta,p)$ instead of its antisymmetric part
$\delta f_a(\theta,p)$, we will always have $\langle \delta f, B \delta f \rangle
\ge \langle \delta f_a, B \delta f_a \rangle$. We then arrive at the following
necessary and sufficient condition of linear stability:
\begin{eqnarray}
\label{necsuff}
\fl
&-& \thpint \frac{1}{\gamma(\theta,p)}((D \delta f)(\theta,p))^2 \\ \fl &+& 
\thpint \int_{-\infty}^{\infty} \dd \, p' \, \int_0^{2\pi} \dd \, \theta' \,
(D \delta f)(\theta,p) V(\theta -\theta') (D \delta f)(\theta',p')
\ge 0 \nonumber
\end{eqnarray}
for any $\delta f(\theta,p)$, with the equality holding only when
$(D \delta f)(\theta,p) = 0$. This is the main expression of this section. In
treating the particular case of the HMF model, it will be the basis to obtain
a condition on the stationary state $f(\veps(\theta,p))$.

\section{The energy principle and the most refined formal stability criterion}
\label{secsympl}

The stability of stationary states of the Vlasov equation has been
studied by Kandrup \cite{kandrup90,kandrup91}, introducing a
Hamiltonian formulation for this equation, obtaining a sufficient
condition of linear stability. We give here few details, mainly to
show that, for stationary states of the form $f(\veps(\theta,p))$,
with $f'(\veps(\theta,p)) < 0$, this condition becomes identical to
the one given in Eq.  (\ref{necsuff}), thus becoming also necessary.

This approach is based on a Hamiltonian formulation of the Vlasov
equation, and on the observation that the Vlasov dynamics admits an
infinite number of conserved quantities, called Casimir invariants,
given by:
\begin{equation}
\label{casimirdef}
C_A[f] = \thpint A(f(\theta,p)) \, ,
\end{equation}
for any function $A(x)$. Note that functionals of the form
(\ref{expfunct}) are particular Casimirs. The Hamiltonian formulation
of the Vlasov equation (\ref{vlasovgen}) for $f(\theta,p)$ is
realized by casting it in the form
\begin{equation}
\label{hamilvlas}
\frac{\partial f}{\partial t} + \left\{f,\veps \right\} = 0 \, ,
\end{equation}
where $\veps$ is the individual energy (\ref{enelocvlas1}) and the
curly brackets denote the Poisson bracket:
\begin{equation}
\label{poissonbra}
\left\{ a,b\right\} = \frac{\partial a}{\partial \theta}\frac{\partial b}{\partial p}
-\frac{\partial a}{\partial p}\frac{\partial b}{\partial \theta} \, .
\end{equation}
It can be shown that, if $f(\theta,p)$ is a stationary state of the
Vlasov equation of a general form, then a sufficient condition for its
linear stability is the following \cite{kandrup91}: the difference
between the total energy (\ref{energgen1}) computed at the perturbed
state $f(\theta,p) +\delta f(\theta,p)$ and the total energy computed
at the stationary state $f(\theta,p)$ is positive for all $\delta
f(\theta,p)$ that conserve all the Casimirs. In other words,
$f(\theta,p)$ is linearly stable if it is a local minimum of energy
with respect to perturbations that conserve all the Casimirs. This
forms the most refined formal stability criterion. These so-called
``phase-preserving'' or symplectic perturbations can be expressed in
the form:
\begin{equation}
\label{symplperturb}
f(\theta,p) + \delta f(\theta,p) = e^{\left\{ a, \cdot \right\}} f(\theta,p) \, ,
\end{equation}
for some ``small'' generating function $a(\theta,p)$. They amount to a
re-arrangement of phase levels by a mere advection in phase
space. Expanding to second order in $a$ we have
\begin{equation}
\label{expandf}
f(\theta,p) + \delta f(\theta,p) = f(\theta,p) + \left\{a,f\right\} +
\frac{1}{2}\left\{a,\left\{a,f\right\}\right\} \, .
\end{equation}
The corresponding expansion of the total energy (\ref{energgen1}) is easily obtained as:
\begin{eqnarray}
\label{expandenertot}
\fl
&&E[f + \delta f] - E[f] = \thpint \veps(\theta,p) \left\{a,f\right\}(\theta,p) \nonumber \\
\fl &+&\frac{1}{2}\thpint \veps(\theta,p)\left\{a,\left\{a,f\right\}\right\}(\theta,p) \nonumber \\
\fl &+& \frac{1}{2} \thpint \int_{-\infty}^{\infty} \dd \, p' \, \int_0^{2\pi} \dd \, \theta' \,
\left\{a,f\right\}(\theta,p)V(\theta - \theta')\left\{a,f\right\}(\theta',p')  \, ,
\end{eqnarray}
where the individual energy (\ref{enelocvlas1}) at the stationary
distribution has been used (and for clarity the explicit dependence of
the Poisson brackets has been written). Now, it is possible to exploit
the identity:
\begin{equation}
\label{identpois}
\thpint c_1\left\{c_2,c_3\right\} = \thpint c_2\left\{c_3,c_1\right\}
\end{equation}
for any three functions $c_1,c_2,c_3$. Using in addition that for a
stationary state $\left\{f,\veps\right\} = 0$, we have that the first
line of the right-hand side of Eq.  (\ref{expandenertot}), i.e., the
first order variation of the total energy, vanishes.  This shows that
the total energy at a stationary distribution is an extremum ($\delta
E=0$) with respect to symplectic perturbations. However, this does not
guarantee that it is an extremum with respect to all
perturbations. Using again the identity (\ref{identpois}), the second order
variations of energy deduced from Eq. (\ref{expandenertot}) are:
\begin{eqnarray}
\label{expandenertot1}
\fl
&&\delta^2 E =
\frac{1}{2}\thpint \left\{\veps,a\right\}(\theta,p)\left\{a,f\right\}(\theta,p) \nonumber \\
\fl &+& \frac{1}{2} \thpint \int_{-\infty}^{\infty} \dd \, p' \, \int_0^{2\pi} \dd \, \theta' \,
\left\{a,f\right\}(\theta,p)V(\theta - \theta')\left\{a,f\right\}(\theta',p')  \, .
\end{eqnarray}
The positive definiteness of this expression is a sufficient condition of linear stability. 

We now suppose that the stationary distribution function is a function
of $\veps(\theta,p)$.  In this case $\left\{a,f\right\} =
f'(\veps)\left\{a,\veps\right\}$. Furthermore, we also have
$\left\{a,\veps\right\} = D a$, where $D$ is the linear differential
operator defined in Eq. (\ref{defD}). Using that $D \veps = 0$,
and therefore $D f'(\veps) = 0$, Eq. (\ref{expandenertot1}) becomes in
this case:
\begin{eqnarray}
\label{expandenertot2}
&&\delta^2 E =
-\frac{1}{2}\thpint \frac{1}{f'(\veps(\theta,p))}((D \tilde{a})(\theta,p))^2 \\
&+& \frac{1}{2} \thpint \int_{-\infty}^{\infty} \dd \, p' \, \int_0^{2\pi} \dd \, \theta' \,
(D \tilde{a})(\theta,p) V(\theta - \theta') (D \tilde{a})(\theta',p') \nonumber  \, ,
\end{eqnarray}
where $\tilde{a} \equiv f'(\veps)a$. The positive definiteness of this
expression is exactly the necessary and sufficient condition of
linear stability (\ref{necsuff}), recalling the definition of
$\gamma(\theta,p)$.

It is convenient to introduce the notation $\delta f(\theta,p)\equiv D\tilde{a}(\theta,p)$.
Then, the necessary and sufficient condition of
linear stability can be written
\begin{eqnarray}
\label{finres}
&-& \thpint \frac{1}{\gamma(\theta,p)} (\delta f(\theta,p))^2 \\ &+& 
\thpint \int_{-\infty}^{\infty} \dd \, p' \, \int_0^{2\pi} \dd \, \theta' \,
\delta f(\theta,p) V(\theta - \theta') \delta f(\theta',p') \ge 0, \nonumber
\end{eqnarray}
for any perturbation of the form $\delta f(\theta,p)\equiv
D\tilde{a}(\theta,p)$ where $\tilde{a}(\theta,p)$ is any
function. These perturbations correspond to a mere displacement (by
the advective operator $D$) of the phase levels, i.e. to dynamically
accessible perturbations. It is straightforward to check by a direct
calculation that these perturbations conserve energy and all the
Casimirs at first order (this is of course obvious for symplectic
perturbations). Indeed, using $\delta f(\theta,p)=\left\{a,f\right\}$
and identity (\ref{identpois}), we get
\begin{eqnarray}
\label{econs}
\delta E &=& \thpint \delta f(\theta,p) \veps= \thpint \left\{a,f\right\} \veps
\nonumber \\ &=& \thpint \left\{f,\veps\right\} a=0,
\end{eqnarray}
and
\begin{eqnarray}
\label{cascons}
\delta C_A &=& \thpint \delta f(\theta,p) A'(f)= \thpint \left\{a,f\right\} A'(f)
\nonumber \\ &=& \thpint \left\{f,A'(f)\right\} a=0.
\end{eqnarray}

\section{Less refined formal stability criteria: sufficient conditions of stability}
\label{formaltot}

As we have seen above, the minimization of $E$ with respect to
symplectic perturbations, i.e. dynamically accessible perturbations
that conserve all the Casimirs, is a necessary and sufficient
condition of linear stability. It is also the most refined criterion
of formal stability since all the constraints of the Vlasov equation
are taken into account individually. Less refined formal stability
criteria, that provide only sufficient (albeit simpler) conditions of
linear stability can be obtained by relaxing some constraints.

\subsection{The ``microcanonical'' formal stability}
\label{microcanonstab}

As we have seen in section \ref{formalgeneral}, a stationary state of
the form $f(\theta,p) = f(\veps(\theta,p))$ with $f'(\veps)< 0$ is
obtained by extremizing a functional $S$ of the form (\ref{expfunct}),
with the function $C(x)$ given by Eq. (\ref{fextremexam}), under the
constraints given by the normalization $I$, Eq. (\ref{normexam}), and
the total energy $E$, Eq. (\ref{energgen1}); but also by extremizing
the total energy $E$ at constant $S$ and $I$. Furthermore, we have
proven that a maximum of $S$ at fixed $E$ and $I$ is a minimum of $E$
at fixed $S$ and $I$ and viceversa. Since we have seen that for
distributions of the form $f(\veps(\theta,p))$ the minimization of $E$
with respect to symplectic perturbations (i.e. perturbations that
conserve all the Casimirs) is a necessary and sufficient condition of
linear stability, it is clear that the minimization of $E$ with
respect to all perturbations that conserve $S$ and $I$ (i.e., two
particular Casimirs) gives a sufficient condition of linear
stability. In turn, this means that the maximization of $S$ at
constant $E$ and $I$ gives the same sufficient condition.

Specializing Eq. (\ref{variatEc}) to our case, we immediately obtain our sufficient
condition:
\begin{eqnarray}
\label{variatsec2b}
&-& \thpint \frac{1}{\gamma(\theta,p)} (\delta f(\theta,p))^2 \\ &+& 
\thpint \int_{-\infty}^{\infty} \dd \, p' \, \int_0^{2\pi} \dd \, \theta' \,
\delta f(\theta,p) V(\theta - \theta') \delta f(\theta',p') \ge 0 \nonumber
\end{eqnarray}
for all $\delta f(\theta,p)$ that at first order give $\delta S =
\delta I = 0$, or equivalently $\delta E = \delta I = 0$.

Another way to see that Eq. (\ref{variatsec2b}) gives a sufficient
condition, if Eq. (\ref{finres}) gives a necessary and sufficient
condition, is the following.  Considering that $D\veps = 0$, we have
that all functions of the type $\delta f=D\tilde{a}$ give $\delta E =
\delta I = 0$ at first order.  Therefore, we arrive at the conclusion
that the condition for the maximization of $S$ at constant $E$ and
$I$, or for the minimization of $E$ at constant $S$ and $I$, is
stronger than the condition for linear dynamical stability. Said
differently, if inequality (\ref{variatsec2b}) is satisfied for all
perturbations $\delta f(\theta,p)$ that conserve $E$ and $I$ at first
order, it is {\it a fortiori} satisfied for all perturbations that
conserve $E$, $I$ and all the Casimirs at first order. However, the
reciprocal is wrong. Therefore, Eq. (\ref{variatsec2b}) gives only a
sufficient condition of linear stability.

At this stage, it is interesting to note some analogies with
thermodynamics.  In particular, the formal stability obtained by
maximizing $S$ at constant $E$ and $I$ can be interpreted as a
``microcanonical'' formal stability problem if we regard $S$ as a
``pseudo entropy''. Less refined stability properties can be found by
relaxing one or both constraints (see below). 

On the other hand, taking $S$ as being the Boltzmann entropy, we note
that thermodynamical stability (in the usual sense) implies Vlasov
linear dynamical stability. However, the converse may not be true in
the general case, i.e. the Maxwell-Boltzmann distribution could be
linearly stable according to (\ref{finres}) without being a maximum
of Boltzmann entropy at fixed energy and normalization (i.e. a
thermodynamical state).

\subsection{The ``canonical'' formal stability}
\label{canonstab}

Following the usual procedure of thermodynamics, we pass from the
``microcanonical'' problem of maximizing $S$, Eq. (\ref{expfunct}), at
constant $E$ and $I$, to the ``canonical'' problem of maximizing the
``pseudo free energy'' $S -\beta E$ (equivalent to minimizing $E
-\frac{1}{\beta}S$) at constant $I$. Introducing the Lagrange
multiplier $\mu$, we obtain again the first order variational problem:
\begin{equation}
\label{extremfirstbis}
\delta S - \beta \delta E - \mu \delta I = 0 \, ,
\end{equation}
equivalent to Eq. (\ref{extremfirst}), and therefore the same
extremizing stationary state. Without repeating again the computations
made in section \ref{formalgeneral}, it is now clear that the
condition of maximum (i.e., of formal stability) is given by the
relation
\begin{eqnarray}
\label{variatsec2c}
&-& \thpint \frac{1}{\gamma(\theta,p)} (\delta f(\theta,p))^2 \\ &+& 
\thpint \int_{-\infty}^{\infty} \dd \, p' \, \int_0^{2\pi} \dd \, \theta' \,
\delta f(\theta,p) V(\theta - \theta') \delta f(\theta',p') \ge 0 \nonumber
\end{eqnarray}
for all $\delta f(\theta,p)$ that at first order give $\delta I = 0$.

\subsection{The ``grand-canonical'' formal stability}
\label{grandcanonstab}

Relaxing also the constraint of normalization is associated to the
passage to the ``grand-canonical'' problem. Namely, we look for the
maximum of the ``pseudo grand-potential'' $S - \beta E -\mu I$ (or the
minimum of $E-\frac{1}{\beta}S + \frac{\mu}{\beta}I$) without any
constraint. The first order variational problem will be again given by
Eq. (\ref{extremfirstbis}), thus obtaining the same stationary state,
while the condition of formal stability is given by the relation
\begin{eqnarray}
\label{variatsec2d}
&-& \thpint \frac{1}{\gamma(\theta,p)} (\delta f(\theta,p))^2 \\ &+& 
\thpint \int_{-\infty}^{\infty} \dd \, p' \, \int_0^{2\pi} \dd \, \theta' \,
\delta f(\theta,p) V(\theta - \theta') \delta f(\theta',p') \ge 0 \nonumber
\end{eqnarray}
for all $\delta f(\theta,p)$.

This unconstrained problem corresponds to the usual energy-Casimir
method \cite{holmmars}.

\subsection{Summary of stability problems}

We have found that the necessary and sufficient condition of
linear dynamical stability for a stationary state $f(\veps(\theta,p))$
of the Vlasov equation is given by Eq. (\ref{necsuff}). We have proven
that this is equivalent to the fact that the stationary distribution
function $f$ satisfies (locally) the problem:
\begin{equation}
\label{condkand}
\min_f \left\{ E[f] \ |\ \,\,\, {\rm all}\,\, {\rm Casimirs} \right\}.
\end{equation}
This is the most refined criterion of formal stability as it takes
into account an infinity of constraints. By relaxing some constraints,
we have then found that progressively less refined, sufficient
conditions of linear stability are given by the following
problems. First, the ``microcanonical'' stability problem:
\begin{equation}
\label{maxmicro} 
\max_f \left\{ S \ | \,\,\, E, I \right\} \, ,
\end{equation}
equivalent to:
\begin{equation}
\label{maxmicro2} 
\min_f \left\{ E \ | \,\,\, S, I \right\} \, .
\end{equation}
Then, the ``canonical'' stability problem:
\begin{equation}
\label{maxcanon} 
\max_f \left\{ S -\beta E \ | \,\,\, I \right\} \, .
\end{equation}
Finally, the ``grand-canonical'' stability problem:
\begin{equation}
\label{maxgrandcanon} 
\max_f \left\{ S - \beta E -\mu I \right\} \, .
\end{equation}
The solution of an optimization problem is always solution of a more
constrained dual problem \cite{ellisineq}. Therefore, a distribution
function that satisfies the ``grand-canonical'' stability problem (no
constraint) will satisfy the ``canonical'' stability problem, a
distribution that satisfies the ``canonical'' stability problem (one
constraint) will satisfy the ``microcanonical'' stability problem, and
a distribution that satisfies the ``microcanonical'' stability problem
(two constraints) will satisfy the infinitely constrained stability
problem (\ref{condkand}). This is the analogous of what happens in the
study of the stability of macrostates in thermodynamics. Of course,
the converse of these statements is wrong and this is similar to the
notion of ensembles inequivalence in thermodynamics. We have the chain of
implications
\begin{equation}
(\ref{maxgrandcanon})\Rightarrow (\ref{maxcanon})\Rightarrow (\ref{maxmicro2})
\Leftrightarrow (\ref{maxmicro})\Rightarrow (\ref{condkand}).
\end{equation}

The usefulness of these less refined conditions of linear stability
will be clear after the stability conditions, that now appear as
conditions to be satisfied by the perturbation to the stationary
distribution function, will be transformed, in the application to the
HMF model, in explicit conditions on the stationary distribution
function itself. It will be shown that the less refined conditions of
stability are associated to simpler expressions, and therefore, in a
concrete calculation, one might use the simpler expressions if the
more refined ones appear to be practically unfeasible. The procedure
is to start by the simplest problem and progressively consider more
and more refined stability problems so as to prove (if necessary) the
stability of a larger and larger class of distributions. Of course, if
we can prove the stability of all the distribution functions with a
particular criterion (see, e.g., Appendix), it is not necessary to
consider more refined criteria.

{\it Remark:} The connection between the optimization problems
(\ref{condkand})-(\ref{maxgrandcanon}) was first discussed in relation
to the Vlasov equation in \cite{aaantonov}, in Sec. 8.4 of \cite{cd},
and in Sec. 3.1 of \cite{nyquistgrav}. Similar
results are obtained in 2D fluid mechanics for the Euler-Poisson
system \cite{stabvortex}. Criterion (\ref{condkand}) is equivalent to
the so-called Kelvin-Arnol'd energy principle, criterion
(\ref{maxgrandcanon}) is equivalent to the standard Casimir-energy method
 introduced by Arnol'd  \cite{arnold} and criterion
(\ref{maxmicro}) is equivalent to the refined stability criterion given by
Ellis {\it et al.} \cite{eht}.

\section{The linear dynamical stability of Vlasov stationary states of the HMF model}
\label{stabhmftot}

For the HMF model we have $V(\theta - \theta') = -\cos (\theta - \theta')$. The
extremization of a functional of the type (\ref{expfunct}) leads to
a function of the type (see Eq. (\ref{fextremexam})):
\begin{equation}
\label{fextremhmf}
f(\theta,p) = F \left[\beta \left( \frac{p^2}{2} - M_x(f) \cos \theta
- M_y(f) \sin \theta \right) + \mu \right] \, ,
\end{equation}
with $\beta > 0$, and with $M_x(f)$ and $M_y(f)$ given by self-consistency
equations
\begin{equation}
\label{selfconsx}
M_x(f) = \int_{-\infty}^{\infty} \dd \, p' \, \int_0^{2\pi} \dd \, \theta' \,
\cos \theta' f(\theta',p')
\end{equation}
and
\begin{equation}
\label{selfconsy}
M_y(f) = \int_{-\infty}^{\infty} \dd \, p' \, \int_0^{2\pi} \dd \, \theta' \,
\sin \theta' f(\theta',p') \, .
\end{equation}
These are the two components of the magnetization.  In this case the
mean field potential $\Phi(\theta;f)$ is
\begin{equation}
\label{meanfieldpothmf}
\Phi(\theta;f) = -M_x(f) \cos \theta -M_y(f) \sin \theta \, ,
\end{equation}
and therefore the individual energy is given by:
\begin{equation}
\label{eneparthmf}
\veps(\theta,p) \equiv \frac{p^2}{2} + \Phi(\theta;f)= \frac{p^2}{2}
- M_x(f) \cos \theta - M_y(f) \sin \theta \, .
\end{equation}
Substituting Eqs. (\ref{meanfieldpothmf}), (\ref{selfconsx}) and (\ref{selfconsy}) in
Eq. (\ref{energgen1}), we obtain the total energy:
\begin{equation}
\label{energhmf}
E = \thpint \frac{p^2}{2} f(\theta,p) \,\, -\frac{1}{2}
(M_x^2(f) + M_y^2(f)) \, .
\end{equation}
The Vlasov equation for the HMF model reads:
\begin{equation}
\label{vlasovhmf}
\frac{\partial f(\theta,p)}{\partial t} + p \frac{\partial f(\theta,p)}{\partial \theta}
-\left(M_x(f) \sin \theta - M_y(f) \cos \theta\right)\frac{\partial f(\theta,p)}{\partial p} = 0 \, .
\end{equation}
For the HMF model, we are interested in studying the stability of stationary
solutions of Eq. (\ref{vlasovhmf}) given by functions of the form
$f(\theta,p) = f(\frac{p^2}{2} - M_x(f) \cos \theta - M_y(f) \sin \theta)$.

Without loss of generality we can suppose that $M_y(f) = 0$ and therefore that
\begin{equation}
\label{fextremhmf1}
f(\theta,p) = F \left[\beta \left( \frac{p^2}{2} - M \cos \theta \right)
+ \mu \right] \, ,
\end{equation}
where for simplicity we have denoted $M \equiv M_x(f)$ (dropping the explicit
dependence on $f$). In this case we have:
\begin{equation}
\label{meanfieldpothmf1}
\Phi(\theta;f) = -M \cos \theta\, .
\end{equation}
The individual energy is:
\begin{equation}
\label{eneparthmf1}
\veps(\theta,p) \equiv \frac{p^2}{2} + \Phi(\theta;f)= \frac{p^2}{2}
- M \cos \theta \, ,
\end{equation}
while the total energy is:
\begin{equation}
\label{energhmf1}
E = \thpint \frac{p^2}{2} f(\theta,p) \,\, -\frac{1}{2} M^2 \, .
\end{equation}

The linearized Vlasov equation, governing the linear dynamics of
$\delta f(\theta,p,t)$ around the stationary distribution
$f(\theta,p)$ of the form (\ref{fextremhmf1}), is obtained by
linearizing Eq. (\ref{vlasovhmf}) (in our case with $M_y(f)=0$ and
$M_x(f)=M$), and it is given by:
\begin{equation}
\label{linearvlasov}
\frac{\partial}{\partial t}\delta f = -p \frac{\partial}{\partial \theta}\delta f
+ M \sin \theta \frac{\partial}{\partial p}\delta f +
p f'(\veps(\theta,p))\frac{\partial}{\partial \theta}\Phi(\theta;\delta f)
\, ,
\end{equation}
where $\Phi(\theta;\delta f)$ must contain also the contribution of $\delta f$
to a magnetization in the $y$ direction:
\begin{equation}
\label{enerdelta}
\Phi(\theta;\delta f) = -\int_{-\infty}^{\infty} \dd \, p' \, \int_0^{2\pi}
\dd \, \theta' \,
\cos (\theta -\theta') \delta f(\theta',p') \, .
\end{equation}
Eq. (\ref{linearvlasov}) is in the form of Eq. (\ref{dynamexam}), with
the linear operator $D$ now taking the form:
\begin{equation}
\label{defDhmf}
(Dg)(\theta,p) = p\frac{\partial}{\partial \theta}g(\theta,p) -
M \sin \theta \frac{\partial}{\partial p}g(\theta,p) \, .
\end{equation}
This operator has the property that $D\veps = 0$ and $D\gamma = 0$,
with $\veps$ given in Eq. (\ref{eneparthmf1}) and where we again use
for simplicity the notation $\gamma(\theta,p)$ for the negative
definite function $f'(\veps)$. The linearized Vlasov equation can then
be written, similarly to Eq. (\ref{dynamexam}), as:
\begin{equation}
\label{dynamhmf}
\frac{\partial \delta f}{\partial t} (\theta,p,t) = -\gamma(\theta,p)
(DK\delta f)(\theta,p,t) \, ,
\end{equation}
where for the HMF model the operator $K$ is defined by:
\begin{equation}
\label{defKhmf}
(Kg)(\theta,p) = \frac{1}{\gamma(\theta,p)} g(\theta,p) -
\int_{-\infty}^{\infty} \dd \, p' \, \int_0^{2\pi} \dd \, \theta' \,
\cos (\theta -\theta') g(\theta',p') \, .
\end{equation}

We are now in the position to follow the general results presented in
section \ref{dynstabgeneral}. Exploiting the particularly simple expression of
the interaction potential in the HMF model we have, from Eq. (\ref{necsuff}),
that the necessary and sufficient condition for the linear stability of
$f(\theta,p)$ is:
\begin{eqnarray}
\label{necsuffhmf}
\fl
&-& \thpint \frac{1}{\gamma(\theta,p)}((D \delta f)(\theta,p))^2 \\ \fl &-& 
\left( \thpint \cos \theta (D \delta f)(\theta,p) \right)^2 -
\left( \thpint \sin \theta (D \delta f)(\theta,p) \right)^2
\ge 0 \, . \nonumber
\end{eqnarray}
We note that in Eqs. (\ref{necsuffhmf}) the first term in the left-hand side is
positive definite, while the second and third terms are negative definite.

We have to find in which case the condition in Eq. (\ref{necsuffhmf}) is satisfied.
We can exploit the antisymmetry of the operator $D$, that implies that the functional subspace
orthogonal to the kernel of the operator is transformed in itself. In fact, if $g_1$ belongs
to the kernel of $D$, and $g_2$ is orthogonal to $g_1$, then:
\begin{equation}
\label{orthoanti}
\langle g_1, Dg_2\rangle = - \langle Dg_1, g_2\rangle = 0 \, .
\end{equation}
The kernel is made of the functions which depend on $(\theta,p)$ through
$\left(\frac{p^2}{2} - M \cos \theta\right)$. We may therefore transform the problem
(\ref{necsuffhmf}) in the problem of satisfying the relation
\begin{eqnarray}
\label{add}
\fl
&-& \thpint \frac{1}{\gamma(\theta,p)}(\delta f(\theta,p))^2 \\ \fl &-& 
\left( \thpint \cos \theta \delta f(\theta,p) \right)^2 -
\left( \thpint \sin \theta \delta f(\theta,p) \right)^2
\ge 0  \nonumber
\end{eqnarray}
subject to the conditions:
\begin{equation}
\label{constrdynam}
\thpint \left(\frac{p^2}{2} - M \cos \theta\right)^s \delta f(\theta,p) = 0
\,\,\,\,\,\,\,\,\, s= 0,1,\dots \, .
\end{equation}
However, we should take into account the case in which the stationary distribution
function $f$ has a power law decay for large $p$. We therefore substitute the previous
conditions with:
\begin{equation}
\label{constrdynamb}
\fl
\thpint h\left[\frac{p^2}{2} - M \cos \theta\right]
\left(\frac{p^2}{2} - M \cos \theta\right)^s \delta f(\theta,p) = 0
\,\,\,\,\,\,\,\,\, s= 0,1,\dots \, ,
\end{equation}
where $h$ is a function that assures integrability; it may be chosen, e.g., equal to
$\exp \left[-\left(\frac{p^2}{2} - M\cos \theta \right)\right]$.

Since the function $\gamma(\theta,p)$ is even in $\theta$, it is
useful to separate $\delta f$ in its even and odd parts in $\theta$,
i.e.  $\delta f(\theta,p) = \delta f_e(\theta,p) + \delta
f_o(\theta,p)$. In this way, our problem to satisfy the relation in
Eq. (\ref{add}) subject to the conditions given in
Eq. (\ref{constrdynamb}), is separated in a pair of separate
problems. Precisely, for the even part we have to satisfy:
\begin{equation}
\label{necsuffhmffuncte}
\fl
- \thpint \frac{1}{\gamma(\theta,p)}(\delta f_e(\theta,p))^2 - 
\left( \thpint \cos \theta \delta f_e(\theta,p) \right)^2
\ge 0
\end{equation}
for all $\delta f_e(\theta,p)$ such that the relations
\begin{equation}
\label{constrdynambe}
\fl
\thpint h\left[\frac{p^2}{2} - M \cos \theta\right]
\left(\frac{p^2}{2} - M \cos \theta\right)^s \delta f_e(\theta,p) = 0
\,\,\,\,\,\,\,\,\, s= 0,1,\dots
\end{equation}
are verified. For the odd part we have to satisfy
\begin{equation}
\label{necsuffhmffuncto}
\fl
- \thpint \frac{1}{\gamma(\theta,p)}(\delta f_o(\theta,p))^2 - 
\left( \thpint \sin \theta \delta f_o(\theta,p) \right)^2
\ge 0
\end{equation}
without any condition.

At this point, we note the following things. Firstly, any $\delta f_e$
such that $\cos \theta \delta f_e$ has a vanishing integral will
trivially give a positive value for the left-hand side of
Eq. (\ref{necsuffhmffuncte}), and similarly any $\delta f_o$ such that
$\sin \theta \delta f_o$ has a vanishing integral will trivially give
a positive value for the left-hand side of Eq.
(\ref{necsuffhmffuncto}). Therefore the only problems can come from
function $\delta f_e$ and $\delta f_o$ that do not have these
mentioned properties.  Secondly, since both (\ref{necsuffhmffuncte})
and (\ref{necsuffhmffuncto}) are quadratic functions of $\delta f_e$
and $\delta f_o$, respectively, the sign of the expression is not
changed by the multiplication of $\delta f_e$ or $\delta f_o$ by any
number.  We can therefore study the sign of the left-hand sides of
(\ref{necsuffhmffuncte}) and (\ref{necsuffhmffuncto}) also by imposing
a linear condition on $\delta f_e$ and $\delta f_o$.

Therefore we proceed in the following way. We look for the extremum of the left-hand side
of Eq. (\ref{necsuffhmffuncte}), constrained by the conditions (\ref{constrdynambe}), with
the further convenient constraint:
\begin{equation}
\label{cond1}
\thpint \cos \theta \delta f_e(\theta,p) = 1 \, .
\end{equation}
Similarly, we look for the extremum of the left-hand side of Eq. (\ref{necsuffhmffuncto})
under the constraint:
\begin{equation}
\label{cond2}
\thpint \sin \theta \delta f_o(\theta,p) = 1 \, .
\end{equation}

It is useful at this point to introduce the following definitions:
\begin{eqnarray}
\label{notatepsalphabis}
\alpha^{(h)}_s &\equiv& \thpint h\left[\frac{p^2}{2} - M\cos \theta \right]
\gamma(\theta,p)\left(\frac{p^2}{2} - M\cos \theta \right)^s \nonumber \\ &=&
\thpint h\left[\frac{p^2}{2} - M\cos \theta \right]\gamma(\theta,p)\left(\veps(\theta,p;f)\right)^s
\end{eqnarray}
and
\begin{eqnarray}
\label{notatepsetabis}
\eta^{(h)}_s &\equiv& \thpint h\left[\frac{p^2}{2} - M\cos \theta \right]
\gamma(\theta,p)\cos \theta \left(\frac{p^2}{2} - M\cos \theta \right)^s
\nonumber \\ &=&
\thpint h\left[\frac{p^2}{2} - M\cos \theta \right]
\gamma(\theta,p)\cos \theta \left(\veps(\theta,p;f) \right)^s
\, ,
\end{eqnarray}
where the dependence on the function $h$ is explicitly indicated.

We begin with the problem related to $\delta f_e$. Introducing the Lagrange multipliers
$2\mu_s^{(h)}$ for the constraints (\ref{constrdynambe}) and $2\nu$ for the constraint
(\ref{cond1}), respectively, the conditioned extremum of the left-hand side of
Eq. (\ref{necsuffhmffuncte}) is given by the equation:
\begin{equation}
\label{extremfe}
\fl
-\frac{1}{\gamma(\theta,p)}\delta f_e(\theta,p) -(1 + \nu)\cos \theta -\sum_{s=0}^\infty
\mu_s^{(h)} h\left[\frac{p^2}{2} - M\cos \theta \right]
\left( \frac{p^2}{2} - M \cos \theta \right)^s = 0 \, .
\end{equation}
It is clear that this extremum is a minimum, since the the second variation is
simply $-\frac{1}{\gamma}>0$. Therefore the necessary and sufficient condition
is that the disequality (\ref{necsuffhmffuncte}) is satisfied for the extremal
$\delta f_e(\theta,p)$ determined by Eq. (\ref{extremfe}).
Denoting furthermore $\xi = -(1+\nu)$, Eq. (\ref{extremfe}) gives:
\begin{equation}
\label{extremfe1}
\delta f_e(\theta,p) = \xi \gamma(\theta,p) \cos \theta -\sum_{s=0}^\infty \mu_s^{(h)}
\gamma(\theta,p) h\left[\veps(\theta,p;f)\right]\left(\veps(\theta,p;f)\right)^s \, .
\end{equation}
Substituting this expression in Eqs. (\ref{constrdynambe}),
we obtain the system of equations:
\begin{equation}
\label{constreqs}
\sum_{s'=0}^\infty \mu_{s'}^{(h)} \alpha_{s+s'}^{(h)} = \xi \eta_s^{(h)} \, \, \, \, \, \, \,
\, \, \, s=0,1,\dots \, ,
\end{equation}
while substitution in Eq. (\ref{cond1}) gives:
\begin{equation}
\label{constreqs1}
\xi \thpint \gamma(\theta,p) \cos^2 \theta - \sum_{s=0}^\infty \mu_s^{(h)}
\eta_s^{(h)} = 1 \, .
\end{equation}
From the system (\ref{constreqs}), we may obtain the multipliers $\mu_s^{(h)}$ as a function
of the multiplier $\xi$. We see in particular that the multipliers $\mu_s^{(h)}$ are
proportional to $\xi$. We therefore introduce the ``normalized'' multipliers
$\tilde{\mu}_s^{(h)}$, given by the solution of the system of equations:
\begin{equation}
\label{constreqsnorm}
\sum_{s'=0}^\infty \tilde{\mu}_{s'}^{(h)} \alpha_{s+s'}^{(h)} = \eta_s^{(h)}
\, \, \, \, \, \, \, \, \, \, s=0,1,\dots \, .
\end{equation}
We have that $\mu_s^{(h)} = \xi \tilde{\mu}_s^{(h)}$; substituting in Eq. (\ref{constreqs1})
we obtain:
\begin{equation}
\label{solxi}
\xi = \frac{1}{\thpint \gamma(\theta,p) \cos^2 \theta - \sum_{s=0}^\infty
\tilde{\mu}_s^{(h)} \eta_s^{(h)}} \, .
\end{equation}

The relation (\ref{necsuffhmffuncte}) for $\delta f_e$ equal to the extremal
function given by Eq. (\ref{extremfe1}) can now be easily obtained, taking into account
Eqs. (\ref{cond1}), (\ref{constreqsnorm}) and (\ref{solxi}). Introducing
the further short-hand notation
\begin{equation}
\label{constrshort}
\sum_{s=0}^\infty \tilde{\mu}_s^{(h)} \eta_s^{(h)} \equiv z(\gamma)
\end{equation}
(where we put in evidence the dependence on the stationary
distribution function through $\gamma$), we have:
\begin{equation}
\label{necsuffhmffunctef}
\frac{1}{z(\gamma) -\thpint \gamma(\theta,p) \cos^2 \theta} \ge 1 \, .
\end{equation}
We have thus obtained a relation involving only the stationary
distribution function.  This is the main expression of this paper.

The relation valid in the case of the linear dynamical stability of homogeneous
(i.e., with $M=0$) stationary distribution functions is easily obtained. In fact, in that
case $\eta_s^{(h)} = 0$; therefore $\tilde{\mu}_s^{(h)} = 0$ and thus $z(\gamma) = 0$.
Then, taking into account that the integral of $\cos^2 \theta$ is equal to $\pi$,
Eq. (\ref{necsuffhmffunctef}) becomes in this case:
\begin{equation}
\label{necsuffhmffunctefhomb}
1 +\pi\pint \gamma(p) \ge 0 \, .
\end{equation}
This is identical with the expression generally found in the
literature for the linear stability of homogeneous distribution
functions in the HMF model (see, e.g.,
Ref. \cite{ik,choi,yama,cvb,cd}); for this comparison we have to
consider that for homogeneous distribution functions $\gamma(p) =
f'(p)/p$.  We will show that for homogeneous distribution functions
all extremal problems lead to the same result. This explains why the
same expression is obtained from the ``canonical'' formal stability
problem \cite{yama} (in principle, this approach only gives a sufficient condition
of linear stability, but our present results show that it is in fact sufficient
and necessary).

The result just obtained for the problem associated to the even part
$\delta f_e(\theta,p)$ can be immediately transformed in that for the
problem associated to the odd part $\delta f_o(\theta,p)$. In this
case, the only constraint is Eq. (\ref{cond2}), so no multipliers
$\mu_s^{(h)}$ are present. Then, we easily obtain the further condition,
analogous to Eq. (\ref{necsuffhmffunctef}), that has to be satisfied
by $\gamma$; namely:
\begin{equation}
\label{necsuffhmffunctof}
1 + \thpint \gamma(\theta,p) \sin^2 \theta \ge 0 \, .
\end{equation}
For homogeneous distribution functions this relation becomes equal to
Eq. (\ref{necsuffhmffunctefhomb}).  However, it can be easily shown
that for inhomogeneous distribution functions of the form given in
Eq. (\ref{fextremhmf1}), i.e., when $M$ is strictly
positive, Eq. (\ref{necsuffhmffunctof}) is satisfied as an equality,
independently of the particular form of the function and of the value
of its parameters. In fact:
\begin{eqnarray}
\label{forequal}
\fl
\thpint \gamma(\theta,p) \sin^2 \theta &\equiv& \thpint \sin^2
\theta\frac{\partial F}{\partial \veps}
\nonumber \\ \fl
= \frac{1}{M}\thpint \sin \theta \frac{\partial F}{\partial \theta}
&=& -\frac{1}{M}\thpint \cos \theta F = -1 \, .
\end{eqnarray}
This equality is clearly associated to a $\delta f_o(\theta,p)$ that simply rotates, at first
order, the distribution function (\ref{fextremhmf1}). This shows that, for inhomogeneous
distribution functions (\ref{fextremhmf1}), any odd $\delta f_o(\theta,p)$
will satisfy Eq. (\ref{necsuffhmffuncto}).

Summarizing the results of this section, the distribution function
(\ref{fextremhmf1}) is linearly dynamically stable iff the relation
(\ref{necsuffhmffunctef}) is satisfied. This relation reduces to the
simpler form given in Eq. (\ref{necsuffhmffunctefhomb}) for a homogeneous
(i.e., with $M=0$) distribution function.

As we have shown in the general case, the most refined formal
stability criterion (\ref{condkand}) leads to the same necessary and
sufficient condition.

\section{The formal stability of Vlasov stationary states of the HMF model:
sufficient conditions of stability}
\label{formaltothmf}

\subsection{The ``microcanonical'' formal stability for the HMF model}
\label{microcanonstabhmf}

The problem related to the ``microcanonical'' formal stability is obtained by specializing
to the HMF potential the expressions given in section \ref{microcanonstab}.
We thus obtain:
\begin{eqnarray}
\label{necsuffhmffunct}
\fl
&-& \thpint \frac{1}{\gamma(\theta,p)}(\delta f(\theta,p))^2 \\ \fl &-& 
\left( \thpint \cos \theta \delta f(\theta,p) \right)^2 -
\left( \thpint \sin \theta \delta f(\theta,p) \right)^2
\ge 0  \nonumber
\end{eqnarray}
for all $\delta f(\theta,p)$ such that the constraints of
normalization and of total energy are satisfied at first order. Using
the expression of the total energy for the HMF model, we have that the
allowed $\delta f(\theta,p)$ have to satisfy:
\begin{equation}
\label{variatEhmf}
\delta E = \thpint \left( \frac{p^2}{2} - M \cos \theta \right)
\delta f(\theta,p) = 0 \, ,
\end{equation}
and
\begin{equation}
\label{variatIhmf}
\delta I = \thpint \delta f(\theta,p) = 0 \, .
\end{equation}
We immediately see that this is analogous to the linear stability
problem, with the difference that now we have only the constraints
associated to $s=0$ and $s=1$, and the function $h$ is the constant
unitary function. In fact, the constraints (\ref{variatIhmf}) and
(\ref{variatEhmf}) are nothing more than the constraints
(\ref{constrdynambe}) for $s=0$ and $s=1$, respectively, and in the
case $h=1$. Therefore, the only thing we need before writing down the
result for this case is to adapt the definition of the parameters
$\alpha_s$ and $\eta_s$, and of the multipliers $\tilde{\mu}_s$, to
the present situation. We thus define
\begin{equation}
\label{notatepsalpha}
\alpha_s \equiv \thpint \gamma(\theta,p)\left(\veps(\theta,p;f)\right)^s
\end{equation}
and
\begin{equation}
\label{notatepseta}
\eta_s \equiv \thpint \gamma(\theta,p)\cos \theta\left(\veps(\theta,p;f)\right)^s
\, .
\end{equation}
We therefore introduce the ``normalized'' multipliers
$\tilde{\mu}_s$, given by the solution of the system of equations:
\begin{equation}
\label{constreqsnorm1}
\sum_{s'=0}^1 \tilde{\mu}_{s'} \alpha_{s+s'} = \eta_s
\, \, \, \, \, \, \, \, \, \, s=0,1 \, .
\end{equation}

The condition on $\gamma(\theta,p)$ analogous to Eq. (\ref{necsuffhmffunctef})
can now be immediately written down. It is given by:
\begin{equation}
\label{necsuffhmfef}
\frac{1}{w(\gamma) -\thpint \gamma(\theta,p) \cos^2 \theta} \ge 1 \, ,
\end{equation}
where now the short-hand notation $w(\gamma)$ stands for:
\begin{equation}
\label{constrshort1}
\sum_{s=0}^1 \tilde{\mu}_s \eta_s \equiv w(\gamma).
\end{equation}

Summarizing, the distribution function (\ref{fextremhmf1}) is formally
stable with respect to the ``microcanonical'' criterion iff the
relation (\ref{necsuffhmfef}) is satisfied. This relation reduces to
the simpler form given in Eq. (\ref{necsuffhmffunctefhomb}) for a
homogeneous (i.e., with $M=0$) distribution function, since in that
case $\tilde{\mu}_s = 0$. Thus, dynamical linear stability and formal
stability lead to identical conditions for stationary homogeneous
distribution functions.

Although it is not evident from the two expressions
(\ref{necsuffhmfef}) and (\ref{necsuffhmffunctef}), we know that if
the former relation is satisfied, so is the latter, since we had found
that the necessary and sufficient condition for the formal stability
is also a sufficient condition for the linear dynamical stability.

\subsection{The ``canonical'' formal stability for the HMF model}
\label{canonstabhmf}

At this point, it is straighforward to derive the relation for the
less refined formal stability problems. In particular, the ``canonical''
formal stability condition is completely analogous to the
``microcanonical'' formal stability condition, with the difference
that the allowed $\delta f(\theta,p)$ in Eq. (\ref{necsuffhmffunct})
have to satisfy only the normalization constraint, i.e.,
Eq. (\ref{variatIhmf}).  This corresponds to taking only the
constraint associated to $s=0$. In particular, the system
(\ref{constreqsnorm1}) reduces to the single equation:
\begin{equation}
\label{constreqsnorm2}
\tilde{\mu}_0 \alpha_0 = \eta_0 \, .
\end{equation}
Then, the condition on $\gamma(\theta,p)$ now becomes:
\begin{equation}
\label{necsuffhmfefc}
\frac{1}{\frac{\eta_0^2}{\alpha_0} -\thpint \gamma(\theta,p) \cos^2 \theta} \ge 1 \, ,
\end{equation}
or, more explicitly,
\begin{equation}
\label{necsuffhmfefcbis}
\frac{1}{\frac{(\thpint \gamma(\theta,p) \cos \theta )^2}{\thpint \gamma(\theta,p)}
-\thpint \gamma(\theta,p) \cos^2 \theta} \ge 1 \, .
\end{equation}

Again, for $M=0$ this condition becomes identical to Eq. (\ref{necsuffhmffunctefhomb}),
since in that case $\eta_0 = 0$.

For consistency, it is interesting to note that the equality in Eq. (\ref{necsuffhmfefcbis})
corresponds to the condition of marginal stability found by another method
in the Appendix F of \cite{cd}.

\subsection{The ``grand-canonical'' formal stability for the HMF model}
\label{grandcanonstabhmf}

Finally, the ``grand-canonical'' formal stability condition has no
constraint at all. Therefore, the condition on $\gamma(\theta,p)$ is simply:
\begin{equation}
\label{necsuffhmfefd}
-\frac{1}{\thpint \gamma(\theta,p) \cos^2 \theta} \ge 1 \, .
\end{equation}
Again, for $M=0$ this condition becomes identical to Eq. (\ref{necsuffhmffunctefhomb}).

We remark the following point. The problem associated to the
``grand-canonical'' formal stability, Eq. (\ref{maxgrandcanon}), does
not constraint the value of the normalization $I$, that therefore can
be different from $1$. It is clear however, that if we want to study
the ``grand-canonical'' formal stability of a distribution function
which is an extremum also for the other stability problems, we have to
restrict ourselves only to normalized distribution functions.

In the Appendix, just as a useful exercise, we show that, as we should
expect, the inhomogeneous Maxwell-Boltzmann distribution function is
``canonically'' formally stable, and therefore also
``microcanonically'' formally stable and linearly dynamically stable,
but it is not ``grand-canonically'' stable. This shows the role of the
constraints and the importance of considering sufficiently refined
stability criteria.

\section{Discussion and conclusions}
\label{discuss}

In this paper we have derived a necessary and sufficient condition of
 linear stability for a stationary state of the Vlasov
equation. This condition is expressed by
Eq. (\ref{necsuffhmffunctef}), which is the core of the paper. Less
refined conditions of formal stability, which are sufficient,
although not necessary, for linear dynamical stability, have also been
obtained.

We should make two remarks. Firstly, it is clear that the form of the
HMF interaction potential has simplified the task, and that further
computations should be made for more complicated
potentials\footnote[7]{Interestingly, in stellar dynamics, using the Antonov
criterion or the energy principle, it can be shown that all
spherical galaxies with $f=f(\veps)$ and $f'(\veps)<0$ are linearly
\cite{sygnet}, and even nonlinearly \cite{lemou}, stable.}. Secondly,
actual computations for linear dynamical stability will always require
a degree of approximation, since the infinite sum implicit in the
system (\ref{constreqsnorm}) and in the definition of $z(\gamma)$,
Eq. (\ref{constrshort}), will have to be replaced by some finite
representation. This has led us to treat also the less refined formal
stability conditions. At the price to have only sufficiency, more
manageable expressions are to be expected.

We would like to conclude with some comments about the relevance of the Vlasov
stable stationary states from the point of view of thermodynamics.

If the system is initially in a state that is not a stationary state
of the Vlasov equation, it can be argued that there will be a rather
fast evolution until a stable stationary state is reached. However,
some care must be exercised about the sense in which this statement
has to be taken. Analogously to the Liouville theorem for the $N$-body
distribution function of Hamiltonian systems, the time evolution of
the one-body distribution function as governed by the Vlasov equation
is such that its phase levels are conserved (in fact, the Vlasov
equation states exactly the equality to zero of the convective
derivative of $f$). In particular, an initial two levels $f$, i.e. an
$f$ which is constant in a given region of the $(\theta,p)$ plane and
zero outside of this region, will be two levels for all the following
evolution. How can we expect such a function to evolve towards a
smooth stable stationary state characterized by a continuity of phase
values? This can be realized only in a coarse-grained sense, when we
study a sort of smeared one-body distribution function, in which the
value of $f$ at each point is substituted by the average of $f$ taken
in a small neighbourhood of the given point. If there is an efficient
mixing of the dynamics, we may expect that, no matter how small is the
averaging neighbourhood (provided it is not vanishing), the averaged
$f$ will evolve towards the stable stationary state of the Vlasov
equation. This is exactly the framework in which the Lynden-Bell
theory of violent relaxation has been proposed \cite{lynd1967}.

It is a typical reasoning in thermodynamics or statistical mechanics
to argue that the distribution functions of a system, in particular
the one-body distribution function, will evolve according to the
maximization of a functional given some constraints. For example, the
final Boltzmann-Gibbs state of the one-body distribution function will
be given by the maximization of the Boltzmann entropy
\begin{equation}
\label{entropybg}
S_B[f] = - \thpint f(\theta,p) \ln f(\theta,p)
\end{equation}
subject to the constraints of normalization, Eq. (\ref{normgen}), and
given total energy, Eq. (\ref{energgen}); the potential
$\Phi(\theta;f)$ will have to be determined self-consistently. The use
of the Boltzmann-Gibbs entropy (\ref{entropybg}) is fully justified to
characterize the state reached after the ``collisional'' regime has taken
over; it will be the most mixed state given the constraints. In a
collisionless regime such as the one governed by the Vlasov equation, one can
make the same hypothesis about the evolution towards the most mixed
state given the constraints, but one has to take into account that the
mixing, even if maximally efficient, has to take place without
violating the properties of the Vlasov equation, as the conservation
of the phase levels of the distribution function. In this way, the
Lynden-Bell expression of the entropy is obtained \cite{lynd1967,assise}.

The Lynden-Bell maximization problem gives a coarse-grained
distribution of the form $\overline{f}(\theta,p) =
\overline{f}(\veps(\theta,p))$ with $\overline{f}'(\veps)<0$, i.e. a
particular steady state of the Vlasov equation. As such, it extremizes
a functional of the form
\begin{equation}
\label{functionalgen}
S[\overline{f}] = - \thpint C\left(\overline{f}(\theta,p)\right) \, ,
\end{equation}
at fixed normalization and energy. It can be shown that if this distribution
function is a maximum of $S$ at fixed $I$ and $E$, then it is
Lynden-Bell thermodynamically stable (see \cite{stabvortex,bouchet}
for the 2D Euler equation and Sec. V of \cite{assise} for the Vlasov
equation).  According to the present study, if it is a maximum of $S$ 
at fixed $I$ and $E$, it is also granted to be
linearly dynamically Vlasov stable. More generally, it can be shown that
the coarse-grained distribution function associated with a Lynden-Bell
thermodynamical equilibrium is always dynamically stable (because it
is a minimum of energy with respect to phase preserving perturbations),
even if it is not a maximum of $S$ at fixed $E$ and $I$ (see Sec. 7.8. 
of \cite{stabvortex} for the 2D Euler equation).

On the other hand, since it is not guaranteed that the mixing is
always completely efficient (this is referred to as ``incomplete
relaxation''), one may argue that in cases when it is
not efficient the dynamics will evolve trying to maximize, always in
the coarse-grained sense, other functionals of the form
(\ref{functionalgen}) that are not consistent with Lynden-Bell's
theory (see \cite{thlb} and Sec. XII of
\cite{assise}). This is an essentially phenomenological approach. 
For example, the Tsallis functional is one particular case of such
functionals (that are called generalized $H$-functions
\cite{thlb}). The constraints of normalization and total energy are
always present, and it is immediate to see that, if the functional to
maximize is of the form (\ref{functionalgen}) then the solution will
always be a function of the form $f(\theta,p) = f(\frac{p^2}{2} +
\Phi(\theta;f))$. The problem at hand in this case is to show that a
function of this form obtained by extremizing (\ref{functionalgen}) at
fixed $I$ and $E$ is really a maximum, i.e., the value of
(\ref{functionalgen}) decreases if we perturb $f$ without changing the
values of the constraints. This ``microcanonical'' stability problem
has been studied for the Tsallis distributions in \cite{cc}.

We are thus led to the conclusion
that functions of the form $f(\theta,p) = f(\frac{p^2}{2} +
\Phi(\theta;f))$ are relevant also in a thermodynamical sense
(with respect to the collisionless dynamics). However, we have shown
that, while for homogeneous states formal ``microcanonical'' stability
implies linear dynamical stability and viceversa, for inhomogenous
states only the first implication is true. Thus, there might be
inhomogeneous states which are formally ``microcanonically'' unstable
(therefore not relevant from a thermodynamical point of view), that
nevertheless are dynamically linearly stable.

\section*{Appendix: The stability of the Maxwell-Boltzmann distribution
function in the HMF model}

As an exercise we can apply the results of section \ref{formaltothmf}
to the magnetized Maxwell-Boltzmann distribution function, that
realizes the Boltzmann-Gibbs global equilibrium. For this case all the
quantities can be obtained analytically, basically because in this
case $\gamma = -\beta f$.  We can consider from the beginning the
``canonical'' formal stability problem, given by
Eq. (\ref{necsuffhmfefc}). It is immediate to obtain in this case
that $\alpha_0 = -\beta$ and $\eta_0 = -\beta M$. We also have:
\begin{equation}
\label{gammacos2bg}
\thpint \gamma(\theta,p) \cos^2 \theta = -\beta \left( M^2 + \Delta M^2 \right) \, ,
\end{equation}
where we have denoted with $\Delta M^2$ the variance of the
magnetization, i.e., the expectation value $\langle \left(\cos \theta
- M\right)^2\rangle$.  We then find that Eq. (\ref{necsuffhmfefc})
reduces to:
\begin{equation}
\label{necsuffhmffunctbg1}
\beta \Delta M^2 \le 1 \, .
\end{equation}
We first note that, in the homogeneous case, the last expression reduces
to the known relation $\frac{1}{2}\beta \le 1$, i.e., $\beta \le
2$. However, it is not difficult to show that, when $\beta > 2$ and $M>0$,
Eq. (\ref{necsuffhmffunctbg1}) is always satisfied, on the basis of
the graphical construction that gives $\beta$ as a function of $M$,
based on the relation
\begin{equation}
\label{mformaxw}
M = \frac{I_1(\beta M)}{I_0(\beta M)} \, ,
\end{equation}
where $I_1$ and $I_0$ are the modified Bessel functions of order $1$ and $0$, respectively.
We then have:
\begin{equation}
\label{proofmaxw}
\beta \Delta M^2 = \beta \frac{\partial}{\partial (\beta M)}\frac{I_1(\beta M)}{I_0(\beta M)}
= \frac{\partial}{\partial M}\frac{I_1(\beta M)}{I_0(\beta M)} < 1 \, ,
\end{equation}
where the last disequality is a consequence of the graphical solution
of Eq.  (\ref{mformaxw}). Then, we have proven the ``canonical''
formal stability, hence the ``microcanonical'' formal stability and
the linear stability. On the contrary, it can be seen that the ``grand-canonical''
formal stability (\ref{necsuffhmfefd}) does not hold for
magnetized states. In fact, from Eq. (\ref{gammacos2bg}), we find that
this stability would require
\begin{equation}
\label{necsuffhmffunctbg2}
\beta M^2 +\beta \Delta M^2 \le 1 \, ,
\end{equation}
that in turn, using Eqs. (\ref{mformaxw}) and (\ref{proofmaxw}), becomes:
\begin{equation}
\label{necsuffhmffunctbg3}
\beta \left[ \left(\frac{I_1(\beta M)}{I_0(\beta M)}\right)^2 +
\frac{\partial}{\partial (\beta M)}\frac{I_1(\beta M)}{I_0(\beta M)}\right] \le 1 \, .
\end{equation}
However, using the results in Appendix B of Ref. \cite{campajphysa},
it is proven that the left-hand side of the last expression, for $M
>0$ is simply equal to $\beta -1$. Since Boltzmann-Gibbs magnetized
states are realized for $\beta > 2$, the disequality cannot be
satisfied. If we consider a nonnormalized
Maxwell-Boltzmann distribution function, and we denote with $A$ its ``mass'',
then the left-hand sides of the last two relations are multiplied by $A$. The
stability condition then becomes a relation between $A$ and $\beta$:
\begin{equation}
\label{necsuffhmffunctbg4}
A\left(\beta -1\right) \le 1 \, .
\end{equation}

Finally, the reader might want to check that the ``microcanonical''
formal stability condition (\ref{necsuffhmfef}), that must be
satisfied since the ``canonical'' one is satisfied, reduces to the
expression:
\begin{equation}
\label{necsuffhmffunctbg}
1-\beta \Delta M^2 + 2 \beta^2 M^2 \Delta M^2 \ge 0 \, .
\end{equation}
Since Eq. (\ref{necsuffhmffunctbg1}) is verified, then
Eq. (\ref{necsuffhmffunctbg}) is a fortiori verified, as it should.

\end{document}